\documentclass[sigconf,screen]{acmart}
\AtBeginDocument{%
  \providecommand\BibTeX{{%
    \normalfont B\kern-0.5em{\scshape i\kern-0.25em b}\kern-0.8em\TeX}}}

\copyrightyear{2024}
\acmYear{2024}
\setcopyright{rightsretained}
\acmConference[CHI '24]{Proceedings of the CHI Conference on Human Factors in Computing Systems}{May 11--16, 2024}{Honolulu, HI, USA}
\acmBooktitle{Proceedings of the CHI Conference on Human Factors in Computing Systems (CHI '24), May 11--16, 2024, Honolulu, HI, USA}
\acmDOI{10.1145/3613904.3642065}
\acmISBN{979-8-4007-0330-0/24/05}

\usepackage{tablefootnote}
\usepackage{multirow}
\usepackage{soul}
\usepackage{xcolor}
\usepackage{lipsum}
\usepackage{xspace}

\definecolor{darkgreen}{rgb}{0,0.5,0}
\definecolor{orange}{rgb}{1,0.5,0}
\definecolor{teal}{rgb}{0,0.5,0.5}
\definecolor{darkpurple}{rgb}{0.5, 0, 0.5}

\definecolor{color-available}{RGB}{0, 148, 68}
\definecolor{color-slightly-affected}{RGB}{251, 176, 64}
\definecolor{color-affected}{RGB}{241, 90, 41}
\definecolor{color-not-available}{RGB}{237, 28, 36}

\newcommand {\rr}[1]{{\color{black}{#1}\normalfont}}

\newcommand{\etal}{et al.\xspace}
\newcommand{\ie}{\emph{i.e.}\xspace}
\newcommand{\eg}{\emph{e.g.}\xspace}

\newcommand {\systemname}{Human I/O\xspace}

\author{Xingyu Bruce Liu}\authornote{Project was conducted when the first author interned at Google Labs.}
\affiliation{%
  \institution{University of California, Los Angeles}
  \city{Los Angeles}
  \state{California}
  \country{USA}}
\email{xingyuliu@ucla.edu}

\author{Jiahao Nick Li}
\affiliation{%
  \institution{University of California, Los Angeles}
  \city{Los Angeles}
  \state{California}
  \country{USA}}
\email{ljhnick@ucla.edu}

\author{David Kim}
\affiliation{%
  \institution{Google Research}
  \city{Zurich}
  \country{SwitzerLand}}
\email{kidavid@google.com}

\author{Xiang `Anthony' Chen}
\affiliation{%
  \institution{University of California, Los Angeles}
  \city{Los Angeles}
  \state{California}
  \country{USA}}
\email{xac@ucla.edu}

\author{Ruofei Du}\authornote{Corresponding author.}
\affiliation{%
  \institution{Google Research}
  \city{San Francisco}
  \state{California}
  \country{USA}}
\email{me@duruofei.com}
\begin{document}

\title[]{\systemname: Towards a Unified Approach to \\ Detecting Situational Impairments}

\begin{abstract}


Situationally Induced Impairments and Disabilities (SIIDs) can significantly hinder user experience in contexts such as poor lighting, noise, and multi-tasking. 
While prior research has introduced algorithms and systems to address these impairments, they predominantly cater to specific tasks or environments and fail to accommodate the diverse and dynamic nature of SIIDs.
We introduce \systemname, a unified approach to detecting a wide range of SIIDs by gauging the availability of human input/output channels.
Leveraging egocentric vision, multimodal sensing and reasoning with large language models, \systemname achieves a 0.22 mean absolute error and a 82\% accuracy in availability prediction across 60 in-the-wild egocentric video recordings in 32 different scenarios.
Furthermore, while the core focus of our work is on the detection of SIIDs rather than the creation of adaptive user interfaces, we showcase the efficacy of our prototype via a user study with 10 participants. 
Findings suggest that \systemname significantly reduces effort and improves user experience in the presence of SIIDs, paving the way for more adaptive and accessible interactive systems in the future.

\end{abstract}

\begin{teaserfigure}
    \centering
  \includegraphics[width=\linewidth]{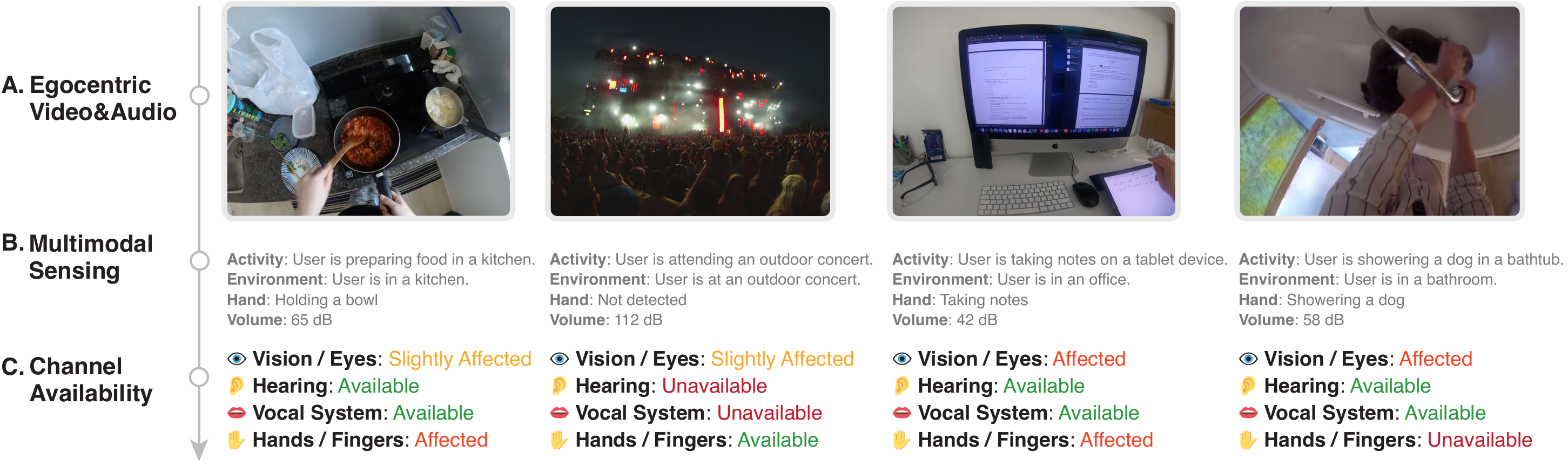}
    \caption{We introduce a new approach to detecting situational impairments based on \textit{the availability of human input/output channels}.
    We instantiate this idea as \systemname, a system that (A) captures egocentric video and audio stream; (B) processes input data and generates a description of the context; and (C) predicts the availability of vision, hearing, vocal, and hands channels.}
    \label{fig:teaser}
\end{teaserfigure}




\maketitle

\section{Introduction}

Everyone experiences Situationally Induced Impairments and Disabilities (SIIDs). These impairments can arise due to various situational factors, such as noise, lighting, temperature, stress, social norms, etc. For example, one might miss an important phone call in a noisy restaurant, or struggle to reply to a text message when doing dishes. These varied situational contexts in daily lives can cause temporary declines in our physical, cognitive, or emotional capacities, leading to unsatisfactory experiences.

Recently, researchers have developed systems to address SIIDs by enhancing situational awareness of mobile devices. 
Most systems employ a ``sense-model-adapt'' design pattern~\cite{wobbrock2019situationally}, that is, to first build a model to identify a particular situation that causes specific SIIDs, and then curate adaptations tailored to that context.
For example, detecting when a person is driving~\cite{dai2010mobile}, walking~\cite{goel2012walktype, kane2008getting}, inebriated~\cite{mariakakis2018drunk}, distracted~\cite{mariakakis2015switchback}, or has rainwater on their touch screen~\cite{tung2018raincheck}.

However, SIIDs are often dynamic and pervasive, making it challenging to scale previous one-off solutions to accommodate users' changing impairments in real-time, across diverse scenarios. 
Consider a typical morning routine: when a person is brushing teeth, they may be constrained from engaging with voice assistants; when washing face, they may struggle with reading urgent messages; and when using a hairdryer, they may miss auditory notifications from their phone. 
Although previous systems have developed models tailored to specific situational impairments, manually designing detection solutions for all possible scenarios and their combinations is impractical and limited in scalability.


In this paper, we propose Human I/O, a new approach that considers SIIDs not as context-specific impairments that require specific detection models, but rather through a unified lens that focuses on the \textit{limited availability of human input/output channels}. 
For instance, rather than devising individual models for activities like face-washing, tooth-brushing, or hair-drying, Human I/O universally assesses the availability of a user's vision, hearing, and hand interaction channels. 
\rr{With the recent development of Large Language Models (LLMs), which exhibit open-vocabulary few-shot learning and reasoning capabilities, we see an exciting opportunity to leverage LLMs and introduce a single, unified framework to identify SIIDs.}
This abstraction broadens our thinking of SIIDs to a comprehensive range of impairments, and allows for the development of an extensible framework that empowers other researchers and developers to continually expand.
Our paper focuses on the comprehensive technical framework to detecting SIIDs, deferring the adaptation of SIID for future research.

We first conducted a formative study with 10 participants to understand the scope of modeling SIIDs based on channel availability. 
These insights emphasize the need for systems to integrate activity, environment, and direct sensing cues for channel availability prediction, and recognize challenges in detecting attentional, affective, and technological SIIDs.
Our findings also suggest that systems should provide varying levels of channel availability, rather than a binary scale as previously assumed in most systems. This will better align with users' needs and allow developers to create tailored strategies based on impairment severity.
We iteratively developed a four-level scale for measuring channel availability: available, slightly affected, affected, and unavailable. 


These insights informed \systemname, a unified system that can automatically detect SIIDs in a wide range of daily activities. 
\systemname leverages (1) an egocentric camera and microphone, (2) computer vision and audio analysis models, and (3) the reasoning capabilities of LLMs to detect SIIDs.
In \systemname's computational pipeline, the system first captures a user's egocentric view with audio and video streams, providing a first-person viewpoint of the user's state. 
The vision and audio models then process the input data, converting it into textual representations. 
\rr{
Finally, we leverage LLMs with chain-of-thought reasoning \cite{wei2022chain} to analyze these textual representations and predict the current availability of human input and ouput channels.
}

We evaluated \systemname on a dataset of 300 clips selected from 60 real-world egocentric video recordings covering 32 distinct scenarios. 
\systemname reaches a 0.22 mean absolute error and 82\% average accuracy on channel availability predictions, with 96\% of predictions deviate by $\leq1$ from actuals.
We also deployed our system in the real world and evaluated it with 10 participants, where they experienced four different scenarios with and without \systemname. Participants found the detection and adaptation for SIIDs significantly reduced their effort level, mental, physical and temporal demands, and improved their user experience.

In summary, we contribute:
\begin{itemize}
    \item A new approach to detecting SIIDs by modeling the availability of human input/output channels.
    \item Insights from a formative study that inform the design of our system, highlighting the need for integrating contextual cues, the scope of our proposed approach, and a four-level scale for measuring channel availability.
    \item The design and implementation of \systemname, which leverages egocentric vision, multimodal sensing, and large language models to predict channel availability across various daily-life situations. \systemname is deployed and open-sourced at \url{https://github.com/google/humanio}.
    \item A technical evaluation of \systemname's performance on a diverse set of 60 in-the-wild egocentric videos, and a user study with 10 participants demonstrating its potential in improving user experience in the presence of SIIDs.
\end{itemize}





\section{Related Work}
Our work builds upon previous research in situationally aware computing, egocentric vision, reasoning by large language models, activity and environmental sensing. 

\subsection{Situationally Aware Computing}
Previous research in human-computer interaction and accessibility have developed systems to model different types of situational impairments. 
A large body of work focused on making mobile devices more situationally aware and capable of improving interaction for users experiencing SIIDs. 
Kane \etal investigated walking user interfaces (WUIs)~\cite{kane2008getting} that adapt their layout based on user movement, demonstrating comparable performance to static interfaces. 
Goel \etal introduced WalkType~\cite{goel2012walktype}, an adaptive text entry system that leverages the mobile device's accelerometer to compensate for movement while walking, improving typing performance. 
In another study, they presented ContextType~\cite{goel2013contexttype}, a system that uses hand posture information to enhance touch screen text entry. 
Mariakakis \etal developed SwitchBack~\cite{mariakakis2015switchback}, a system built upon Focus and Saccade Tracking to help users resume tasks more efficiently in the presence of distractions. They also explored drunk user interfaces~\cite{mariakakis2018drunk}, estimating blood alcohol levels using machine learning models trained on performance metrics and sensor data. 
Tung \etal proposed RainCheck~\cite{tung2018raincheck}, a solution that filters out water-caused touch points on capacitive touchscreens, enhancing interaction accuracy and target selection time.

Although prior art has advanced the field of mobile device usage under various situational impairments, their limitations lie in their narrow focus on specific situations.
These efforts, while significant, only address a fraction of all SIIDs.
Therefore, there remains a need for a more comprehensive approach to detect a broader range of SIIDs. In this paper, we leverage egocentric vision and large language models to address challenges of detecting SIIDs.

\subsection{Egocentric Vision}
The concept of using a wearable camera to gather first-person visual data dates back to the 1970s with Steve Mann's ``Digital Eye Glass'' invention~\cite{mann2005designing}. Since then, wearable cameras have been employed in various health-related applications within the context of Wearable AI. The Microsoft SenseCam uses a lifelogging camera with fisheye lens and trigger sensors, such as accelerometers, heat sensing, and audio devices, to aid those with poor memory as a result of disease or brain trauma~\cite{hodges2006sensecam}.
Kanade and Hebert proposed a prototypical first-person vision system which consists of localization, recognition, and activity recognition components, to provide contextual awareness for caregiving applications~\cite{kanade2012first}. Early computational techniques for egocentric analysis centered on hand-related activity recognition and social interaction analysis, as well as addressing challenges of temporal segmentation~\cite{gunsel1998content} and summarization~\cite{potapov2014category} due to the unconstrained nature of video data. Over the past decade, the field has diversified, with emerging research topics including social saliency estimation~\cite{soo2015social}, privacy-preserving techniques, attention-based activity analysis, hand pose analysis, understanding social dynamics and attention, and activity forecasting~\cite{kitani2012activity}.

In this work, we build upon previous egocentric vision research to develop a more comprehensive approach to situational impairment detection. 
We chose egocentric vision as a means for implementation since it provides the widest ``bandwidth'' of detecting a broad ranges of SIIDs; but the key idea behind our system should be independent of such implementation.

\subsection{Reasoning Capabilities of Large Language Models}

Recent large language models have demonstrated reasoning capabilities via different approaches including zero-shot learning \cite{kojima2022large}, few-shot learning \cite{brown2020language}, chain-of-thoughts \cite{wei2022chain} 
or incorporating multimodal information \cite{zeng2022socratic}. 
These reasoning abilities are particularly useful for tasks such as mathematical problem-solving \cite{lewkowycz2022solving, wei2022chain, frieder2023mathematical}, image-based question answering \cite{zeng2022socratic, li2023blip}, understanding human intents \cite{srinivasan2022quill, Liu2023Visual}, etc.
They have been applied to a broad range of research in HCI community recently including interactive coding support \cite{wang2022enabling, jiang2021genline}, social computing \cite{park2022social}, and communication augmentation \cite{Liu2023Visual}.
For example, Social Simulacra uses LLMs to simulate social interactions and behaviors as social computing prototypes \cite{park2022social}.
Visual Captions leverages a fine-tuned large language model to proactively suggest relevant visuals in open-vocabulary conversations~\cite{Liu2023Visual}. InstructPipe~\cite{Zhou2023InstructPipe} employs a node selector, a code writer, and a code interpreter to create AI pipelines from human instructions.

Such reasoning capabilities of large language models make it possible for our system to, in an open-vocabulary manner, predict the availability of human input/output channels based on the detected activity, environment, and other contextual information.

\subsection{Activity and Environmental Sensing}
A wide variety of sensing technologies and strategies~\cite{lara2012survey,froehlich2009hydrosense,laput2016viband,laput2017synthetic,laput2019sensing,curtiss2021facebit,sun2022human} have been investigated to achieve detection of human activity and measuring a physical environment. 
More relevant to our work are sensing approaches that utilize camera-based \cite{ke2013review,beddiar2020vision} and audio-based \cite{rahman2014bodybeat,laput2018ubicoustics} systems. 
For example, Mo \etal \cite{mo2016human} applied deep learning to classify 5 different locations and 12 distinct activities. 
BodyBeat \cite{rahman2014bodybeat} employs a piezoelectric microphone to detect on-speech body sounds, such as eating noise, breathing, laughter, and coughing. 
Other works \cite{khan2014activity,rodomagoulakis2016multimodal,sun2022human} demonstrate promising outcomes when using multiple modalities.

In our paper, we develop a novel framework that enables the reasoning of human input/output channels through the use of computer vision and audio analysis of video and audio streams. Leveraging large language models, our framework is highly adaptable, allowing for easy integration with both existing and forthcoming sensing technologies for activity and environmental monitoring.





\subsection{SIIDs as the Availability of Human I/O Channels}

\begin{figure}[ht]
\centering
\includegraphics[width=0.7\linewidth]{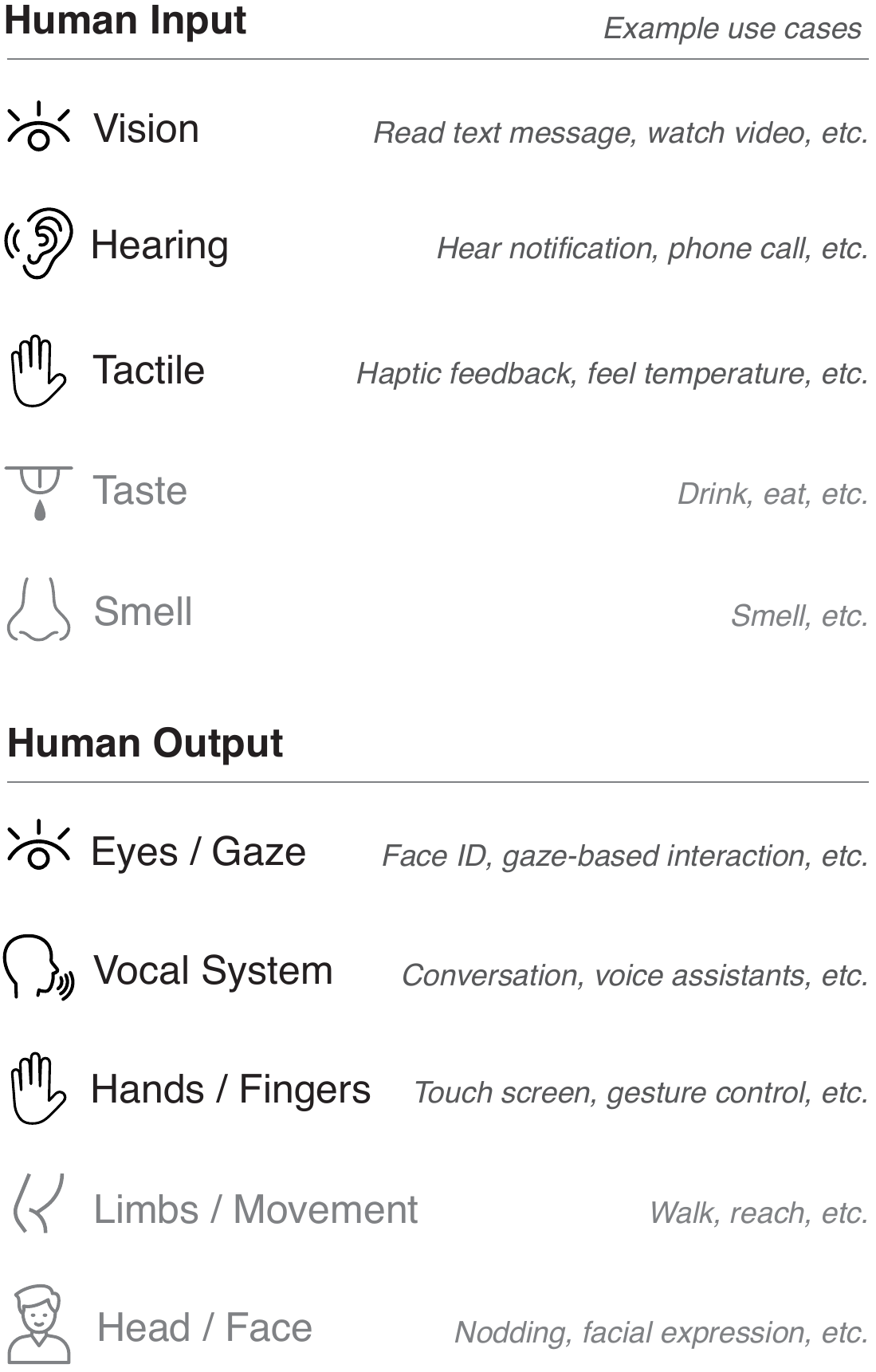}
\caption{Human input/output channels with channels most commonly used in human-computer interaction highlighted in black. We designed and implemented \systemname based on these channels.}
\label{fig:input-output}
\end{figure}

Our motivation for using the availability of human input/output channels for detecting SIIDs lies on Dix et al.'s fundamental model of human-computer interaction~\cite{dix2003human}: humans, similar to computer I/O, receive and send information via different channels. For example, we use vision, hearing, tactile, etc. to receive information coming from the world (input); and use vocal system, eye gaze, hand gestures, etc. to convey information (output). He and Card~\cite{card2018psychology} both describe human as an \textit{``information processing system''} with limited capacity to process information through various channels. 
In addition, Microsoft's Inclusive 101 Guidebook\footnote{Microsoft's Inclusive 101 Guidebook: \url{https://inclusive.microsoft.design}} provides an overview chart of four types of permanent, temporary, and situational impairments: touch, see, hear, and speak, categorized by human sensory channels. Similarly, CrossA11y~\cite{liu2022crossa11y} divides video accessibility issues into lack of information in the visual and auditory channels. 

Building upon these inspirations, we hypothesized that modeling the availability of human input/output channels might provide a more unified approach to detecting SIIDs. 
We summarize a list of human input/output channels from prior work (\autoref{fig:input-output}). In this paper, we focus on channels that are most commonly used in human-computer interaction: vision, hearing, tactile (input), and eyes/gaze, vocal system, hands/fingers (output).  

\section{Formative Study}
To validate the feasibility and further explore the scope of modeling SIIDs as the availability of human input/output channels, we conducted a remote whiteboard session with 10 participants. 
We report on our insights and how they informed our design and implementation of the Human I/O system.

\subsection{Procedure}
\begin{figure}[ht]
\centering
\includegraphics[width=\linewidth]{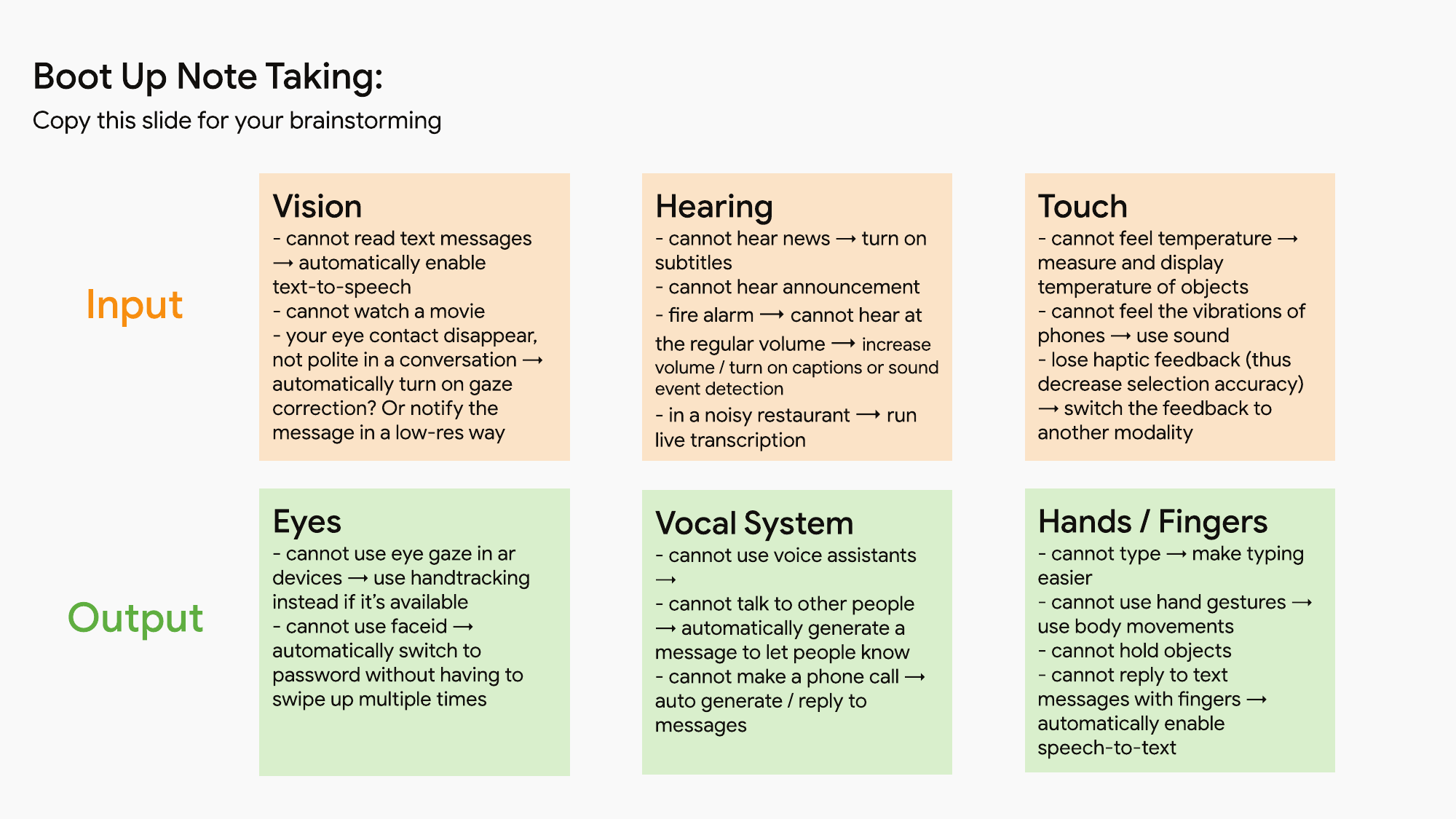}
\caption{An example brainstorming whiteboard from a participant.}
\label{fig:brainstorming}
\end{figure}

We recruited 10 participants via group email invitations and internal communication channels in Google. Participants had various technical and non-technical backgrounds, including software engineers, researchers, UX designers, visual designers, students, etc. 
In a 90-minute online brainstorming session, we first introduced and explained what SIIDs are, showed videos of previous systems that can detect and adapt to SIIDs, and presented our initial ideas on identifying SIIDs by estimating the availability of human input/output channels. 
Participants then brainstormed on a digital whiteboard (\autoref{fig:brainstorming}) based on three prompts: 
(1) For each input/output channel, what are some situations that make it unavailable? (2) For each input/output channel, when it is unavailable, what are some implicated consequences? (3) For each impairment, to what extent would you like to have an adaptive system to intervene versus overcoming it yourself?
We went over each participant's responses and asked them to explain and elaborate on their examples after brainstorming.

\subsection{Findings}


Two researchers organized and analyzed participants' responses with the affinity diagram approach. Informed by the set of themes derived  from the grouped notes, we present insights (\textbf{I1} to \textbf{I5}) around methods to predict channel availability, the scope of our approach in detecting SIIDs, and different levels of channel availability.

\subsubsection{Methods to Predict Channel Availability}
We first asked participants to brainstorm for each channel, what were some situations that would make it unavailable. Participants brainstormed 82 situations in total.
We found that these situations can be broadly detected via three ways (with overlaps) \textbf{(I1)}:

\textit{Activity-based} (46 mentions): Participants identified that sometimes unavailability of a channel is due to the user's engagement in activities, such as driving, cooking, or attending a meeting.

\textit{Environment-based} (26 mentions): Environmental factors were recognized as another source of channel unavailability, such as traveling on an airplane or studying in a library.

\textit{Channel-based} (20 mentions): Participants also noted that some situations could be directly sensed by detecting the state of sensory channels, such as a user wearing headphones, in a loud environment, or holding an object in their hand.

Participants' identified situations align with observations made by Sears et al.~\cite{sears2009physical}, which emphasizes the contribution of both the environment and activity on the existence of impairments and disabilities. 
In addition, direct sensing of channels emerged as a third theme.
By measuring metrics such as the current environmental volume level, or whether the hand is occupied, systems can leverage more direct, lower-level information to predict channel availability.

Furthermore, participants mentioned that some situations can impact more than one channel. For example, playing drums would affect both the user's ability to hear and their hands for interacting with devices.
Certain channels might also be correlated with each other. In particular, the input channel of vision and the output channel of eyes/gaze are almost always available or unavailable together, similarly for tactile and hands/fingers.
In \systemname, we combine human input/output channels into four categories:

\begin{itemize}
    \item Vision / Eye
    \item Hearing
    \item Vocal System
    \item Hands / Fingers
\end{itemize}

\subsubsection{Scope \& Limitations}

We conducted a comprehensive review of the generated situations to understand the boundaries of our approach. Specifically, we sought to identify what types of SIIDs can be effectively modeled by evaluating the availability of human input/output channels and the inherent limitations. 
We compared participants' generated situations to the Situational Factors taxonomy proposed by Wobbrock~\cite{wobbrock2019situationally}, and reviewed what types of situations can be properly detected.  
Wobbrock analyzed a decade's work from 2008 to 2018 on situationally aware mobile devices and categorized different kinds of SIIDs into six categories: \textit{Behavioral}, \textit{Environmental}, \textit{Attentional}, \textit{Affective}, \textit{Social}, and \textit{Technological}. 

Our findings suggest that our approach is capable of identifying situational impairments centered around human sensory abilities, including those induced by \textit{behavioral} (\eg, walking, driving, operating machine), \textit{environmental} (\eg, ambient noise, darkness), and \textit{social} (\eg, conversation, crowd) factors. 
These situations often have a direct impact on the availability of human input/output channels.

However, the approach exhibits limitations in SIIDs that are \textbf{(I2)}: 

\textit{Intrinsic to Human}: Our approach may not be as effective in identifying impairments related to cognitive states, \ie, \textit{attentional} (divided attention, distraction), \textit{affective} factors (stress, fear, fatigue) factors, which may indirectly influence the availability of input/output channels. There is often not a clear mapping between one's mental states and their ability to use input/output channels. 

\textit{Technological}: Impairments induced by technological constraints, like power shortages, weak Wi-Fi connectivity, or hardware limitations—are not directly discernible through the human input/output channel metrics. Such SIIDs will need supplementary techniques or synchronization with device-centric data.




\subsubsection{Levels of Channel Availability}


To understand participants' preferences on how they would like to deal with SIIDs, we also asked them to discuss the extent they would desire system intervention versus overcoming the impairment themselves. This provided insights into 106 impairments, along with their preferred levels of system intervention.
Some interesting examples include: to delay notifications when people's vision channel is currently engaged in activities like driving or biking, generate automatic replies when people's hands are not available to text, and turn on live captions~\cite{Liu2023Modeling} when the hearing channel is not available. 

Participants highlighted that \textit{adaptations may not always be necessary or preferred} \textbf{(I3)},
especially if the unavailability is temporary or can be easily overcome. For instance, when holding a remote controller that occupies the hand, instead of switching to voice input to interact with the device, a user might prefer to put the remote down temporarily and use the hands as the input method again. When taking a sip of coffee, users might not need adaptations for the brief moment when their vocal channel is unavailable.

These observations led us to reconsider, that \textit{channel availability (and SIIDs in general) is not binary} \textbf{(I4)}, as assumed in many previous situationally aware systems. 
We should not simply model a channel as ``available'' or ``unavailable'', ``impaired'' or ``not impaired''. There are many cases in which the channel has a gradient of availability. This could depend on factors such as the difficulty of the ongoing activity, the duration of unavailability, or the user's ability to overcome the situation.

Moreover, \textit{availability of a channel depends not only on the current situation, but also on the incoming task} \textbf{(I5)}. For instance, one participant mentioned, when a user's hands are wet, they might still be usable to perform simple actions such as tapping on the screen to answer an incoming phone call. However, the same hands might be deemed insufficient for more intricate tasks, such as typing to respond to a text message.
This relates back to our previous point that SIIDs are non-binary --- \eg, the wet hand does not cause a binary impairment, but rather exhibits a variable bandwidth in its channel depending on the task at hand.

\subsection{Design Implications}
\label{sec:design-implications}
Based on the insights from our formative study, we outline design implications for systems that models SIIDs based on the availability of human input/output channels:

\noindent(1) \textbf{Consider activity, environment, and direct sensing cues for predicting channel availability (I1):} To more accurately predict channel availability, systems should take into account a combination of activity-based, environment-based, and directly sensing cues. Systems can provide a more comprehensive understanding of the user's situation.
    
\noindent(2) \textbf{Acknowledge the limitations (I2)}: This approach may struggle to identify SIIDs that are attentional, affective, or technological. Designers should be aware of these limitations and consider additional methods for these types of SIIDs.
    
\noindent(3) \textbf{Predict multiple levels of channel availability (I3, I4, I5):} Systems should provide different levels of availability to align users' needs. It is important to note that sometimes users may not want the system to adapt to their situations. Hence, it's essential to provide users with the agency to decide how their SIIDs should be managed. 
This will also allow developers to design different strategies based on the severity.

For \systemname, we developed a four-level channel availability based on insights from our formative study. We randomly selected a total of 20 situations proposed by participants during the session, 5 from each of the vision/eye, hearing, vocal system, and hands/fingers channels.  
Iteratively, two researchers first proposed descriptions of different levels, then coded the 20 situations with the drafted levels, and revised the description of the levels based on the disagreements or ambiguities. 
Researchers repeated this process for three meetings until full agreement was reached.
We identified four levels of channel availability: 

\begin{itemize}
\item \textcolor{color-available}{\textbf{Available}}: The channel is currently not involved in any activity, or constrained by any environmental factors. It takes low to zero effort to use the channel to do a new task. \\ \textit{Example: A user is sitting at their desk with their hands free, eyes not engaged in any task, and no background noise interfering with their hearing or speech.}

\item \textcolor{color-slightly-affected}{\textbf{Slightly Affected}}: The channel is engaged in an activity or constrained by an environmental factor. Given a new task that requires the channel, users can multitask, easily pause and resume to the current activity, or easily overcome the situation. \\ \textit{Example: A user is holding a remote control, which can be quickly put down to free up their hand for another task.}

\item \textcolor{color-affected}{\textbf{Affected}}: The channel is involved in an activity or constrained by an environmental factor. Given a new task, the user may experience inconvenience or require some effort to use the channel. \\ \textit{Example: A user is carrying grocery bags in both hands, making it challenging to use their hands for other tasks without putting the bags down first.}

\item \textcolor{color-not-available}{\textbf{Unavailable}}: The channel is completely unavailable due to an activity or environmental factor, and the user cannot use it for a new task without substantial changes, significant adaptation or changing the environment. \\ \textit{Example: A user is attending a loud concert, making it impossible for them to hear incoming notifications or carry on a conversation without stepping outside.}
\end{itemize}

We observed that the distinction amongst these four levels hinges on the amount of effort for a user to free up a channel for an interactive task and re-occupy the channel later.
To validate the consistency and applicability of these levels, we continued to label all remaining situations independently. We computed the inter-rater reliability and the result shows a high level of agreement between raters, with Cohen's Kappa $\kappa$ = 0.847.

\begin{figure*}[ht]
\centering
\includegraphics[width=0.95\linewidth]{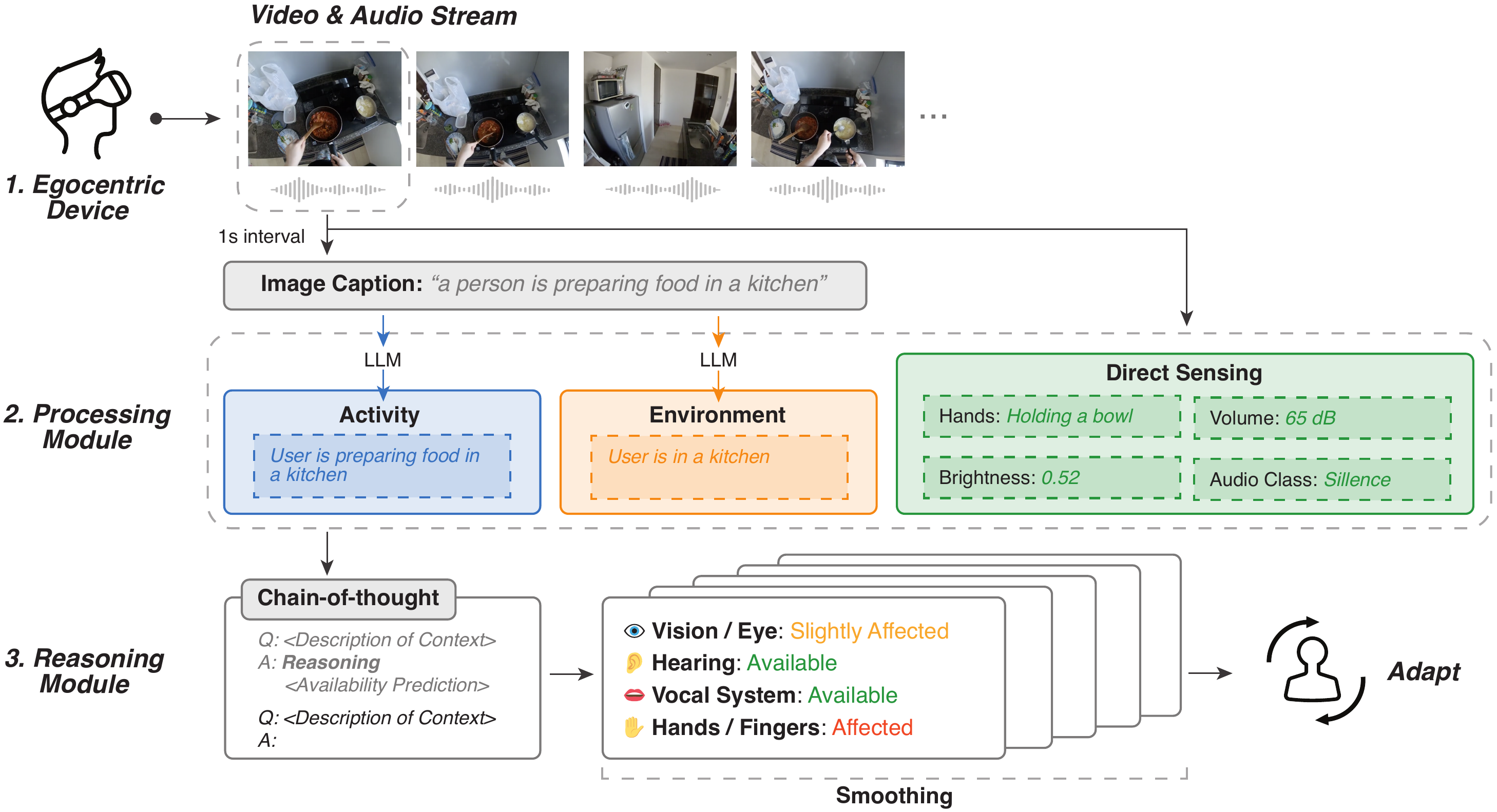}
\caption{The \systemname pipeline comprises three components: (1) an camera and microphone capturing the user's egocentric video and audio stream; (2) video and audio data processing using computer vision, NLP, and audio analysis to obtain contextual information, including user's activity, environment, and direct sensing ; and (3) sending contextual information to a large language model with chain-of-thought prompting techniques, predicting channel availability, and incorporating a smoothing algorithm for enhanced system stability.}
\label{fig:system}
\end{figure*}

\section{Human I/O System}

Following our formative study insights, we developed \systemname, a system that detects situational impairments based on the availability of human input/output channels.

\subsection{Overview}
The \systemname computational pipeline, illustrated in \autoref{fig:system}, consists of three components:
(1) An egocentric camera and microphone capturing video and audio streams of the user's current situation (\autoref{fig:system}.1).
(2) A processing module that processes the video and audio data in one-second intervals using a combination of computer vision, natural language processing, and audio analysis algorithms, and generates a rich set of data, including the user's activity, environment, and direct sensing of specific channels (\autoref{fig:system}.2). 
(3) A reasoning module that leverages a large language model to process the contextual information. It employs chain-of-thought prompting to predict the availability of vision/eye, hearing, vocal, and hands/fingers channels. A smoothing algorithm is incorporated at the end to enhance system reliability (\autoref{fig:system}.3).

We implemented \systemname as a web application to offer a versatile and accessible platform both for users to learn about their SIIDs in daily lives, and developers to debug and evaluate detection methods (\autoref{fig:interface}). 
This approach enables connectivity with different cameras and microphones and allows for easy experimentation on different devices, such as mobile phones, tablets and AR glasses. Furthermore, the web app supports testing on both live video streams and pre-recorded videos, providing a flexible environment for evaluation and user studies. A live demo and open-sourced repository of \systemname can be found at \url{https://github.com/google/humanio}.

\subsection{Data Capture}

In our research setup, \systemname uses a webcam (Logitech C930e) and its integrated microphone to obtain real-time video and audio streams for data capture.
We envision that future implementations of \systemname would seamlessly integrate with lightweight, all-day AR glasses~\cite{olwal2020wearable} equipped with an array of sensors, such as cameras, LiDARs, microphones, eye trackers, and inertial measurement units (IMUs). These sensors will enable richer data capture and provide more comprehensive input to enhance the system's capability.

\begin{figure}[ht]
\centering
\includegraphics[width=0.8\linewidth]{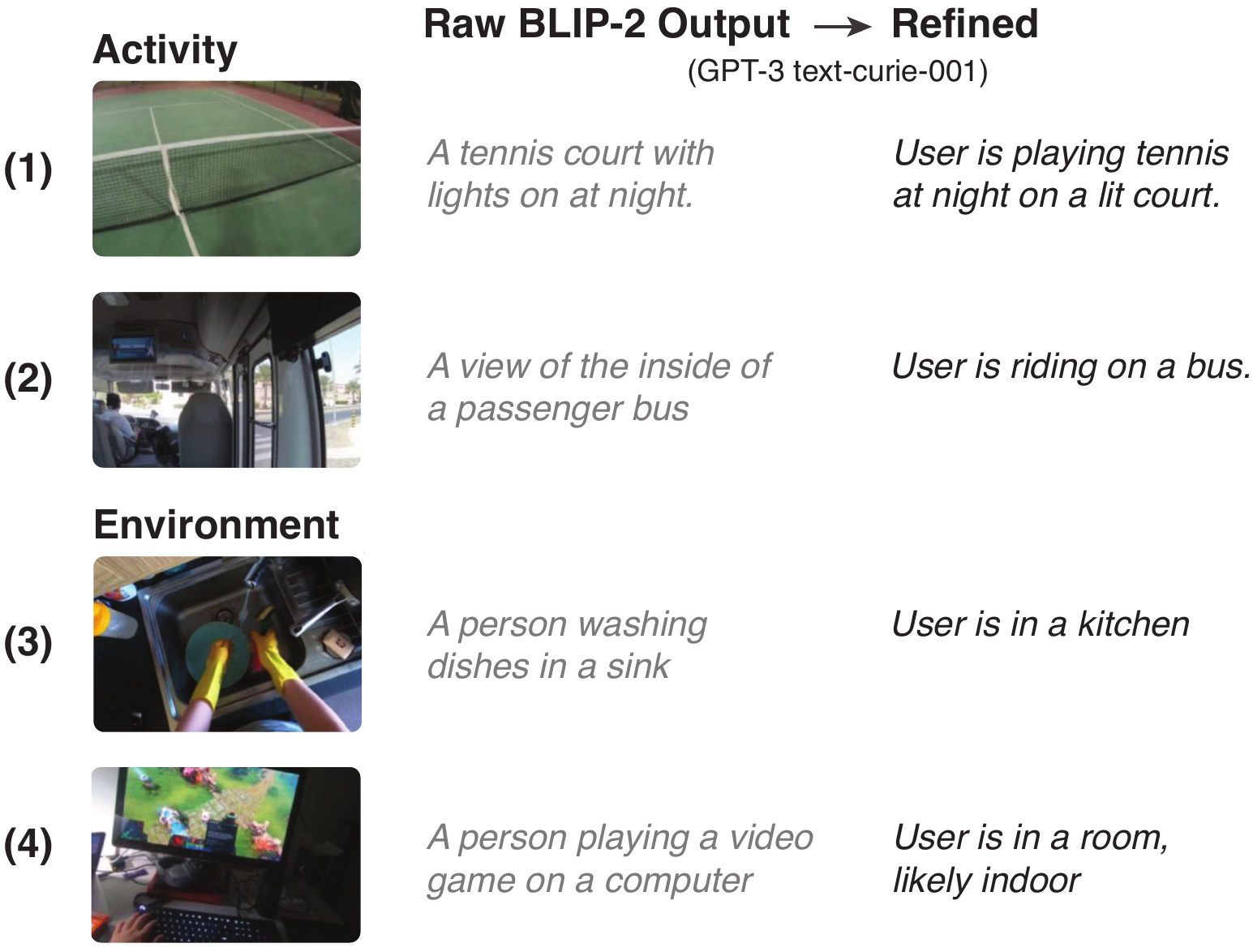}
\caption{Examples of using GPT-3 (text-curie-001) to refine raw image caption results from BLIP-2 to get more accurate descriptions of the current activity and environment.}
\label{fig:activity-env}
\end{figure}

\subsection{Processing Module}
\systemname recognizes the user's context by analyzing video and audio data in one-second intervals.

\subsubsection{Activity Description}
To detect the current activity in an egocentric video, we employ a two-step process. First, we generate an image caption of the current video frame (compressed to 640×480) using the state-of-the-art image captioning model in early 2023, BLIP-2~\cite{li2023blip}.
Although BLIP-2 generates high-quality and objective descriptions, it occasionally falls short in providing an explicit ``activity'' description, \ie, what the person wearing the camera is doing. 
For instance, consider the examples shown in \autoref{fig:activity-env}.

The BLIP-2 output for a tennis court frame is \textit{``a tennis court with lights on at night''}, while for a frame capturing the inside of a bus, it outputs \textit{``a view of the inside of a passenger bus''}. These captions describe the scenes but do not effectively convey the user's actions.

To address this limitation, we integrate the BLIP-2 model with GPT-3 text-curie-001, a faster version of the GPT-3 model capable of simpler tasks. We use the following prompt structure to  guide the GPT-3 model to generate a more accurate activity description (full version in \autoref{sec:prompts}): 

\begin{quote}
   \textit{``An egocentric view of User is showing'' + \\ \texttt{<BLIP-2 output>} + \\ ``Describe what User is doing concisely. Answer in the format of `User is ...'.'' }
\end{quote}

By combining the two models, we obtain a refined activity description that better reflects the user's actions. Referring back to the examples in \autoref{fig:activity-env}.1 and \ref{fig:activity-env}.2, the integrated output for the tennis court frame becomes \textit{``User is playing tennis at night on a lit court''}, while for the bus frame, it produces \textit{``User is riding on a bus''}.

\subsubsection{Environment Description}
To identify user's current environment, we once again combine the outputs of the BLIP-2 and GPT-3 text-curie-001 model. We forward the image caption from BLIP-2 to GPT-3 using the following prompt structure:

\begin{quote}
   \textit{``An egocentric view of User is showing'' + \\ \texttt{<BLIP-2 output>} + \\ ``What location or environment is User likely to be in? Answer in the format of `User is in...'.'' }
\end{quote}

This improves the quality of environment descriptions. For instance, consider an egocentric video frame showing a person washing dishes (\autoref{fig:activity-env}.3). The original BLIP-2 caption states, \textit{``A person washing dishes in a sink''}. With the integration of the GPT-3 model, the output is refined to, \textit{``User is in a kitchen''}. For a video frame displaying a person playing a computer game (\autoref{fig:activity-env}.4), the initial BLIP-2 caption reads, \textit{``A person playing a video game on a computer'}'. Following the integration with GPT-3, the description becomes: \textit{``User is in a room, likely indoors''}.

\subsubsection{Direct Sensing}
Since activity and environment detection may miss information that is not be adequately represented by their high-level descriptions, we also implement direct sensing techniques to gather a more comprehensive set of data. Specifically, we consider hand detection, volume level, audio classification, and environmental brightness.

First, our hand detection algorithm consists of three stages. We first use the MediaPipe Hands model \cite{zhang2020mediapipe} to obtain keypoint localizations of 21 3D hand-knuckle coordinates for both hands. If no hand is detected at this stage, the process is halted, and outputs \textit{``No hand is detected.''}

If a hand is detected, we proceed by utilizing the MediaPipe object detection model (efficientdet\_lite0)  \cite{mpobjectdetection} to detect if either hand is holding any object and what object is it holding (details in \autoref{sec:tech-details}).
If a hand is holding an object, the system outputs \textit{``Hand is holding \texttt{<Object>}.''}

Finally, if a hand is detected but not holding an object, we use the BLIP-2 Visual Question Answering (VQA) model to ask, ``What are the hands doing?''. This unconstrained approach is particularly useful when hand landmarks or objects are not accurately recognized (\eg, object detection model cannot identify drumsticks), or in complex scenarios such as typing, washing hands, etc. 
In this case, we use the result of BLIP-2 VQA as output. 



In addition, \systemname directly senses the availability of the hearing and vision channels through volume level, audio event classification, and brightness measurements. Volume level is determined using the Web Audio API, and measurements are smoothed and converted to decibels. Audio event classification leverages the pre-trained YAMNet model~\cite{yamnet}, capable of detecting 521 distinct audio events. Brightness is assessed using relative luminance, following the WCAG accessibility guideline and Rec. 709 coefficients\footnote{Relative luminance: \url{https://www.w3.org/WAI/GL/wiki/Relative_luminance}}. Temporal smoothing is applied to volume level and brightness measurements. For further details on the implementations, please refer to \autoref{sec:tech-details}.

\subsection{Reasoning Module}
\begin{figure}[ht]
\centering
\includegraphics[width=0.7\linewidth]{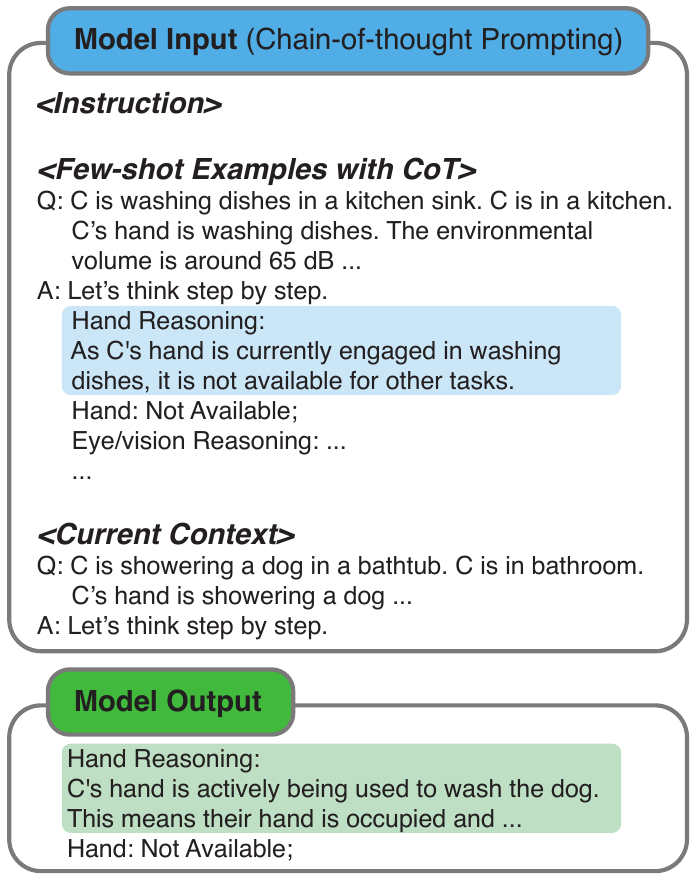}
\caption{An illustration of our prompt structure leveraging chain-of-thought (CoT, highlighted) to enable LLMs to predict channel availability from the context.}
\label{fig:llm}
\end{figure}

We integrate all intermediate results and employs chain-of-thought (CoT) prompting~\cite{wei2022chain} with GPT-4 to predict the availability of human input/output channels.
CoT involves providing the model with a prompt composed of triples: 
\texttt{<input, CoT, output>}, where the chain-of-thought contains a series of intermediate natural language reasoning steps leading to the final output. This method allows the model to accumulate and maintain context and consistency throughout the prediction process.

While CoT is prevalent in LLM research, our method converts multimodal data into structured text, merging audio-visual and language models. Combining textual descriptions with quantitative sensory data, our approach offers richer context, enhancing system accuracy and nuance in predictions.
For \systemname, the prompt consists of three components: \textit{Instruction}, \textit{Few-shot Examples}, and \textit{Current Context}. Please refer to \autoref{fig:llm} for the prompt structure and \autoref{sec:prompts} for the complete prompt used in our system.

\subsubsection{Instruction}
A prefix that clarifies the definitions of our four-level scale availability, as outlined in \autoref{sec:design-implications}, and the task. The task is described as follows:
\begin{quote}
\textit{``Given the current activity and environment as described below, determine the availability of the user's vision/eye, hearing, vision, and hands/fingers channels.''}
\end{quote}

\subsubsection{Few-shot Examples}
Three few-shot examples are provided, each comprising input, chain-of-thought, and output. We selected three examples representing different availability in various channels, different activities and environments: washing dishes in a kitchen, playing an acoustic guitar in a room, and working on a laptop in a library.
For each few-shot example, the input contains activity and environment descriptions, along with direct sensing outputs for hands, volume level, audio classification, and brightness.

We construct the chain-of-thought by defining the intermediate reasoning steps that the model should follow to derive the availability of a channel based on the context. For example, in the case of the hearing channel, the chain of thought may involve considering the volume level, the presence of noise or other sound events, and the user's current activity (\eg playing guitar) and environment (\eg in a library). The model is guided through these reasoning steps and prompted to provide a final output predicting the availability of the hearing channel. A similar process is followed for the other I/O channels.

The output is a four-level availability score, as described in \autoref{sec:design-implications}. 

\subsubsection{Current Context}
We combine the formatted ouputs from the processing module.
For example, for the situation in \autoref{fig:system}, the combined context is:
\begin{quote}
    \textit{Q: User is preparing food in a kitchen. User is in a kitchen. User's hand is holding a bowl. The environmental volume is around 65dB. No audio event is detected in the environment. The luminance value of the current environment is 0.52, in the range of 0 to 1. }
\end{quote}

\begin{figure}[ht]
\centering
\includegraphics[width=\linewidth]{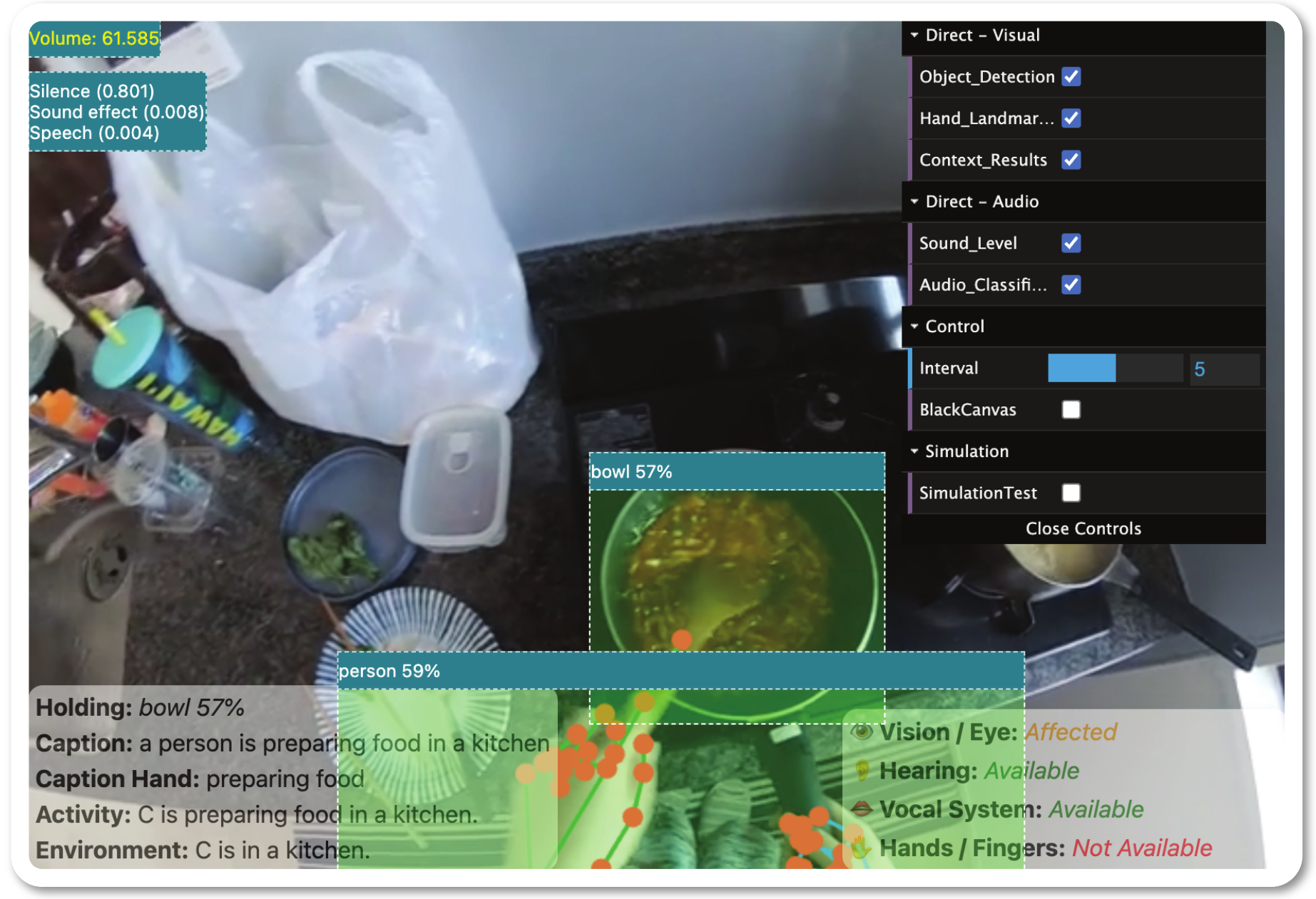}
\caption{The web-based visualization interface of our prototype \systemname system. The prediction results are displayed in the bottom-right. Intermediate auditory information is shown on the top-left corner, and visual information is shown on the bottom-left. The settings is shown as a drop-down menu on the top-right. Detected objects and hands are also highlighted in the video in real-time.}
\label{fig:interface}
\end{figure}

\subsubsection{Availability Prediciton}
We combine instruction, few-shot examples, and the current context as the final input to GPT-4. 
We also added a suffix \textit{``A: Let's think step by step''}~\cite{kojima2022large}.
The model output contains the reasoning and availability prediction of each channel.

\subsubsection{Temporal Smoothing}
To account for potential fluctuations in the prediction output, we apply a smoothing window of size 5 to the model's availability predictions. \systemname runs every second, and with the smoothing window in place, it only outputs an availability prediction if more than three of the past five predictions are the same. This majority prediction is then output as the final availability determination. If there is no majority prediction within the past five predictions, the system outputs ``Unsure''. This smoothing technique reduces the potential impact of brief, sporadic changes in the user's context or sensing outputs.

\subsubsection{Lite Version}
Although chain-of-thought provides robust reasoning to ensure prediction accuracy, generating the reasoning texts is the bottleneck of the system speed.
To enable more real-time prediction, we created a lite version by using GPT-3.5-turbo and removing all intermediate reasoning steps in the few-shot examples in the prompt and \textit{``Let's think step by step''} (\autoref{sec:prompts}). This substantially decreases inference speed to under 1 second, as the system only needs to generate availability predictions for four channels (around 10 tokens), with a lighter model. However, we observe a decrease in performance. We provide a quantitative analysis comparing the two approaches in \autoref{sec:tech-eval}.

\subsection{\systemname Visualization Interface}
We developed a \systemname Visualization Interface (\autoref{fig:interface}), \rr{primarily as a tool for developers and researchers}, to provide real-time monitoring of the processing outputs (along with hand landmarks and object detection) and predictions of availability for all four communication channels. 
The visualization interface has been designed with the flexibility to analyze both live and pre-recorded video feeds. We can conveniently conduct technical evaluations by loading pre-recorded videos and analyzing the data.
In addition, the interface offers a logging system that records all intermediate results throughout the processing pipeline. 
The interface is deployed live at \url{https://github.com/google/humanio}.

\section{Technical Evaluation with In-the-wild Videos}
\label{sec:tech-eval}
We evaluated \systemname on 300 clips from a set of 60 in-the-wild egocentric video recordings under 32 different scenarios.
We report the accuracy (mean absolute error and classification accuracy) and the consistency (intra-video variance) of our system on channel availability predictions.

\begin{figure}[ht]
\centering
\includegraphics[width=0.85\linewidth]{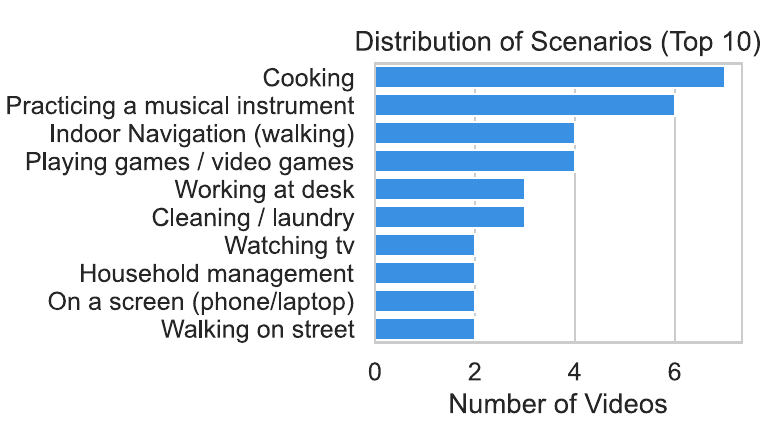}
\caption{Distribution of the top 10 most common scenarios in our evaluation dataset. Full distribution of the scenarios is shown in Appendix \autoref{sec:appendix-technical-eval}.}
\label{fig:scenarios}
\end{figure}

\subsection{Materials}
Our sample was sourced from Ego4D v1~\cite{grauman2022ego4d}, an extensive egocentric dataset with over 3,670 hours of daily-life activity videos, recorded in-the-wild. We applied a filter to select videos shorter than two minutes with audio, and then randomly selected 60 videos, ensuring a diverse range of scenarios. 
Each video features a single, coherent activity.
For each video, we randomly selected 5 non-overlapping clips aligned with the main activity, totaling 300 tested clips.
Our final sample comprises 32 distinct scenarios (\autoref{fig:scenarios}), such as cooking, practicing a musical instrument, working, playing with pets, cleaning, doing laundry, walking, sitting on a sofa etc.

We executed \systemname on each video, logging various data points, including image caption, activity description, environment description, direct sensing results (hand presence, audio volume, audio class, and brightness), LLM reasoning output, and channel availability predictions. 
We recorded the smoothed outputs over the video windows. To compare performance, we also ran the lite version of \systemname.

\subsection{Data Annotation}
Two researchers first independently watched the videos, and annotated the level of availability for vision/eyes, hearing, vocal system and hands/fingers channels based on the four-level scale we developed (\autoref{sec:design-implications}) on a spreadsheet.
Similar to our formative study results, we reached a high level of inter-rater agreement (Cohen's Kappa $\kappa$ = 0.862). We then discussed the mismatches and resolved all disagreements. 
In the final sample, 10.0\% of the channels are labeled as ``Unavailable'', 18.33\% ``Affected'', 30.42\% ``Slightly Affected'', and 41.25\% ``Available''.

\subsection{Metrics}
We assess \systemname using three quantitative metrics:
(1) \textit{Mean Absolute Error (MAE)}: This metric captures the average discrepancy between predicted availability levels and the actual ground truth for each of the four channels within every video clip. Given the nature of our data, MAE offers a precise error depiction. This is because misclassifying a level as ``Slightly Affected'' instead of ``Affected'' is less erroneous than confusing ``Unavailable'' with ``Available''. Availability levels are numerically translated to values ranging from 1 to 4.
(2) \textit{Averaged Classification Accuracy (ACC)}: In addition to MAE, we provide an accuracy measurement for a more intuitive understanding of classification performance.
(3) \textit{Intra-video Variance (VAR)}: Assessing the model's consistency throughout a continuous audio-visual stream is crucial to prevent flickering predictions. The variance score is determined by first calculating it based on video ID and channel then averaging the results.

\begin{table}[ht]
\begin{tabular}{@{}lrrrrrr@{}}
\toprule
               & \multicolumn{3}{c}{\textbf{Human I/O}}                                   & \multicolumn{3}{c}{\textbf{Human I/O Lite}}                     \\ \cmidrule(l){2-7} 
Channels       & \multicolumn{1}{l}{\textit{MAE}} & \multicolumn{1}{l}{\textit{ACC}} & \multicolumn{1}{l}{\textit{VAR}} & \multicolumn{1}{l}{\textit{MAE}} & \multicolumn{1}{l}{\textit{ACC}} & \multicolumn{1}{l}{\textit{VAR}} \\ \midrule
Vision/Eyes   & 0.25 & 76.0\% & 0.17 & 0.45 & 62.0\% & 0.16 \\
Hearing        & 0.23 & 86.7\% & 0.05 & 0.37 & 63.3\% & 0.11 \\
Vocal System  & 0.08 & 92.0\% & 0.01 & 0.23 & 87.3\% & 0.20 \\
Hands/Fingers & 0.33 & 72.7\% & 0.18 & 0.70 & 46.7\% & 0.30 \\ \midrule
\textbf{Total} &\textbf{ 0.22} & \textbf{81.8\%} & \textbf{0.10} & 0.44 & 64.8\% & 0.19 \\ \bottomrule
\end{tabular}
\caption{Technical evaluation of \systemname and \systemname Lite. 
We report the mean absolute error (MAE), average classification accuracy (ACC), and average intra-video variance (VAR) for four channels and the overall results. 
Our system estimates availability levels with small margins of error and variance. 
In \systemname, 96.0\% of predictions are within a discrepancy of 1 from the actual value.}
\label{tab:tech-eval}
\end{table}

\subsection{Results}
We analyzed the performance of \systemname in predicting the availability of each input/output channel. Table \ref{tab:tech-eval} summarizes our results. The system achieves an mean absolute error (MAE) of 0.22, and an average accuracy of 81.8\% across all channels.
\rr{
All MAEs of the four channels are under 0.33. These low MAE values indicate that the system's predictions closely align with the actual availability, with deviations being less than a third of the actual level on average.
Breaking down the performance by individual channels, we observed accuracies of 76.0\% for eyes, 86.7\% for hearing, 92.0\% for vocal, and 72.7\% for hands, with corresponding MAEs of 0.25, 0.23, 0.08, and 0.33. Notably, 96\% of the system's predictions differed from the actual values by 1 or less, and no predictions had a difference greater than 2, demonstrating a reliable performance across different channels.
Additionally, the overall low variance in predictions (0.1), with channel-specific variances of 0.17 (eyes), 0.05 (hearing), 0.01 (vocal), and 0.18 (hands), reflects the system's consistency. This consistency is important for practical applications, as it ensures that the system reliably predict the level of impairments across different instances in the recording.
}

For \systemname Lite, we observe overall slightly inferior performance compared to the full model. However, the MAE for \systemname Lite is still at a very low level around 0.44, showing promising ability to predict SIIDs even with reduced computational resources.

\subsection{Latency}
We evaluated our computation pipeline's latency: \systemname with a full chain-of-thought prompt had an average latency of 19.95 seconds with GPT-4, and 7.33 seconds with GPT-3.5-turbo.
The \systemname Lite (without chain-of-thought reasoning) exhibited a significant improvement, with an average latency  of 2.07 seconds. 

From our observations in the dataset, user's channel availability typically spans more than one minute. This suggests that even a delayed prediction could still be accurate when using it in real-time.
In addition, while latency might matter more for shorter activities that last only a few seconds, our formative study suggests that those short activities likely do not necessitate adaptations.
However, we acknowledge that certain scenarios may still demand low latency. For instance, when users require rapid adaptations in response to changing situations (\eg, entering a noisy subway while still on an important phone call).

Future system can incorporate temporal segmentation techniques to detect changes in scenes or activities, thus eliminating the need for the system to run continuously at one-second intervals. \rr{Alternatively, as the current system's main bottleneck is the inference speed of LLMs, the use of lighter weight models, potentially fine-tuned on an extensive dataset, could be employed to enable faster computation and improve system performance.
Such models may also provide a more concise representation of SIIDs predictions.}
\rr{
In addition, future work could explore identifying pre-impairment scenarios. That is, to anticipate situational impairments before they occur, thereby preparing the system for a timely response. For instance, the system might start making inferences as soon as a faucet is opened, predicting the imminent unavailability of hands due to washing.
While acknowledging the potential for false positives, this method may enhance responsiveness of the system particularly in dynamic environments.
}

\subsection{Failure Cases Analysis}

Our results indicate that the system have similarly effective performance at predicting the availability of the vision, hearing and vocal channels, while the performance in predicting hand availability is less satisfactory. After reviewing the failure cases, we speculate that this lower performance for the hands channel may be attributed to several factors: (1) The complexity of hand-related tasks and the wide range of possible hand impairments make it challenging to accurately capture all nuances of hand-related SIIDs. (2) Occlusion that might affect the quality of the data captured by the egocentric vision system. (3) The few-shot examples may not be sufficiently diverse to represent hand-related SIIDs, thus affecting the model's performance in this specific channel.

In addition, we observed that many incorrect predictions are related to inaccurate activity recognition. The system tends to fail in situations with unclear activities, such as walking around in the bathroom. However, it performs well when the activities are more explicit, such as washing hands, cooking food, or playing drums. 
Future versions of the system can explore balancing the weights for the more confident direct sensing results, and reduce the over-reliance on high-level descriptions of the activity and environment. 

The system also struggles to discern subtle differences between similar activities. For instance, consider two kitchen-related videos where one person is washing their hands and another is stirring food in a pan. The system detects both activities as ``preparing food in the kitchen'', and classifies hand availability as ``unavailable'' in both cases. However, hands can be briefly used for other tasks while stirring food, so they should be considered as ``affected'' rather than ``unavailable''.

Anecdotally, even when the system misrecognizes the activity completely (\textit{e.g.}, scrubbing wood with sandpaper as climbing a wooden ladder), it sometimes still provides relatively accurate channel predictions due to similarities in hand occupancy, environmental volume level, and other factors. 

\section{Experiencing \systemname in Real-time}
\label{sec:user-study}

While our focus remains on the detection of SIIDs, we conducted an additional study to understand users' integrated experiences when \systemname is employed to enable common interface adaptation strategies.




\subsection{Procedure}

We conducted a user study with \systemname Lite to understand its potential and challenges for real users in assisting them in managing incoming tasks during various SIIDs, when combined with common adaptation strategies. The study setup involved an egocentric camera (Logitech C930e), AR glasses (Nreal Light) for adaptive displays, and a custom website displayed on a touch screen tablet to simulate incoming tasks. The website was connected to the \systemname system via a web socket.

We recruited 10 participants from Google (age range: 22--36, avg=28.0, std=3.95) with diverse backgrounds, including students, software engineers, research scientists, designers, product managers, and marketing analysts. Half of the participants reported familiarity with AR wearables (rating > 3 on a scale of 1 to 5).

Participants first familiarized themselves with \systemname. They then simulated four scenarios in a lab space: working, washing hands, hair drying, and eating. Scenarios were selected to represent impairments in each of the vision, hands, hearing, and vocal channels. 
During each scenario, participants received a notification or a task on the tablet, including phone calls, text messages, and video watching. 
We designed corresponding adaptations strategies on \systemname to provide adaptations when impairments were detected. For example, if the hands channel is detected to be affected, \systemname will automatically display a phone call on AR glasses and prompt users to answer/reject the call by saying ``yes'' or ``no''; if the hearing channel is affected, it will automatically turn on captions for a video. Details of all user study scenarios are shown in \autoref{tab:simulation}.
Each scenario was conducted with and without \systemname assistance in a within-subject design with counter-balanced order.
\rr{All participants experienced four scenarios with and without \systemname. Participants followed the adaptation strategies to overcome the situational impairments when \systemname is activated. When \systemname is not active, they were asked to address the incoming task and resume back to their previous task (\eg, pause hand washing and answer the phone call).}

\begin{table*}[ht]
\begin{tabular}{@{}lllll@{}}
\toprule
\# & \textbf{Scenario} & \textbf{Incoming Task} & \textbf{Impaired Channel} & \textbf{Adaptation}         \\ \midrule
1  & Working (Typing)  & Receive a text message & Vision/Eyes               & Display message  on glasses \\
2  & Hand Washing      & Receive a phone call   & Hands/Fingers             & Switch to voice commands    \\
3 & Hair Drying  & Watch a video        & Hearing & Turn on captions   \\
4 & Eating Chips & Receive a phone call & Vocal   & Suggest auto-reply \\ \bottomrule
\end{tabular}
\caption{Scenarios, incoming tasks, and corresponding adaptions participants experimented with in our simulation user study.}
\label{tab:simulation}
\end{table*}


After each scenario, participants completed the NASA Task Load Index (TLX) questionnaire, assessing mental demand, physical demand, temporal demand, overall performance, effort, and frustration level on a 7-point scale (from 1--Lowest to 7--Highest). We also conducted a semi-structured interview to gather qualitative feedback on the participants' experiences.

\begin{figure}[ht]
\centering
\includegraphics[width=0.9\linewidth]{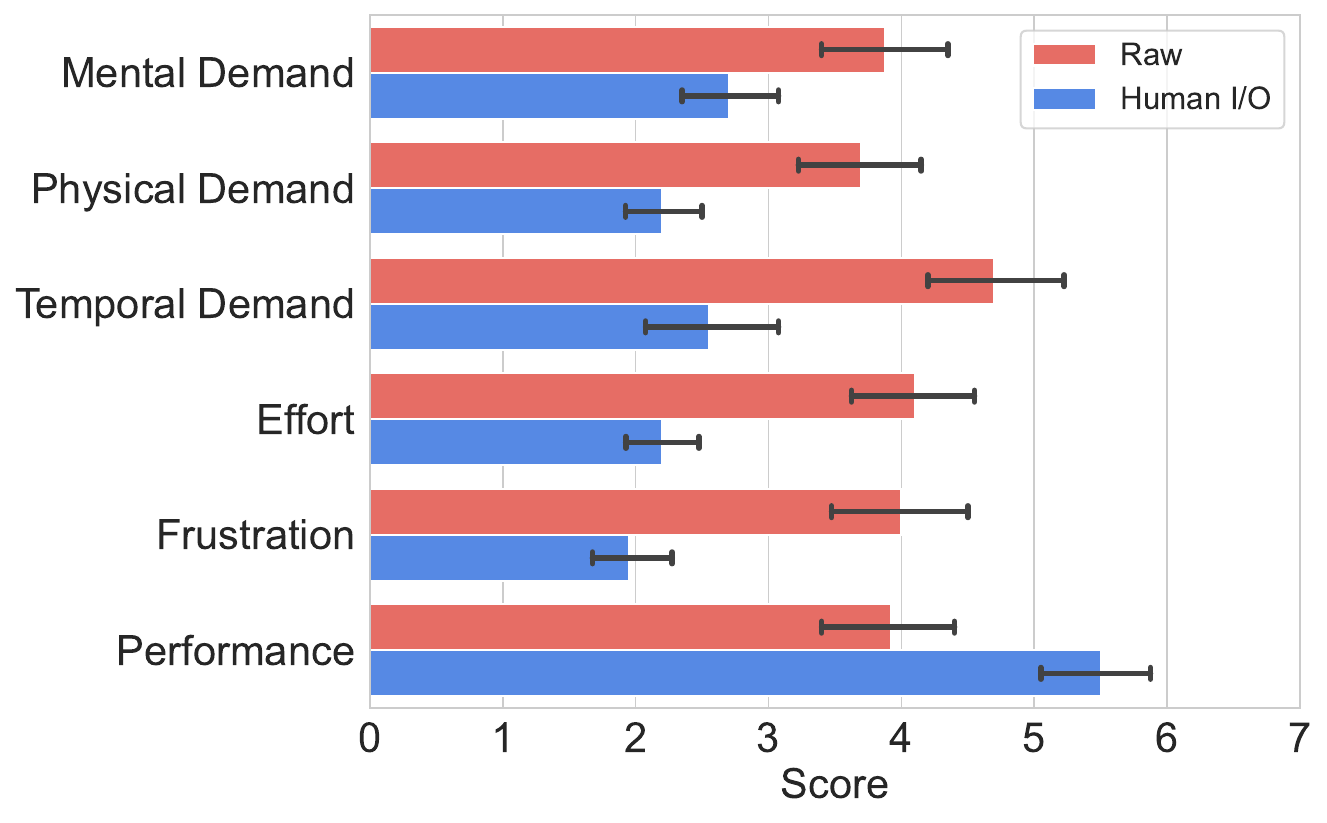}
\caption{Participants’ ratings to Task Load Index questions (on a scale of 1-low to 7-high) for their experience with SIIDs with and without \systemname in user study. All rating differences are statistically significant with \(p<0.001\) via Wilcoxon signed-rank tests.}
\label{fig:taskload}
\end{figure}

\subsection{Findings}


\noindent\textbf{\systemname Enhances User Experience.}
Participants unanimously preferred the \systemname experience. They found the system to be robust (P8, P9), accurate (P1, P2, P3, P6, P10), and and helpful (P7, P8, P9) in various aspects.
Participants reported that \systemname significantly reduced their mental, physical, and temporal demand, as well as their effort and frustration level, and significantly improved their performance to manage incoming tasks during SIIDs (\autoref{fig:taskload}, all \(p < 0.001\) via Wilcoxon signed-rank tests). 
Participants mentioned that when not using the system, they often had to \textit{``delay my physical activities or my intent to interact''}. 
Participants highlighted the system's capability to support their workflow without interruption, as it allowed them to continue their current activities without delays. P4 noted that \systemname enabled them to \textit{``maintain [their] existing workflow and focus on the task at hand.''} This sentiment was echoed by P6, who appreciated the ``time-saving'' aspect of the system.
P7 also emphasized the convenience and efficiency that \systemname provided:
\begin{quote}
    \textit{``It was much more convenient. Especially when my hand is captured by something else, then I'm still able to respond promptly to the request. It's essentially allowing me to do the thing that I want to do but easier and way faster.''} --- P7
\end{quote}

\noindent\textbf{Need for Personalization.}
Interestingly, we observed that for some tasks the perceived usefulness of \systemname varied among participants.
In particular, while some users found \systemname less helpful for tasks that involved quickly interacting with phone notifications and then returning to work, others reported that the notifications severely interrupted their focus and thought processes (P2). The impairments in these cases are not severe enough to warrant the adaptations.

As another example, 6 participants mentioned that the handwashing scenario was the most helpful because \textit{``my hands are all wet and there's practically no way to touch the screen.''} (P4).
However, P5 did not share this sentiment, stating that they were comfortable interacting with their device: \textit{``If my I just wash my hands with water I think it's acceptable for me to click the screen''}.

In light of these varied responses, we see a need for a more personalized availability scale into \systemname to better address individual preferences. 
Researchers and developers should note that it might be wrong to assume that users always want the UI to adapt based on their situations. 
Recognizing this difference in user preference underscores the importance of integrating different levels and options within \systemname.

Participants also expressed desire in customizing the adaptation procedures, emphasizing that familiarity with these adaptations would make the system much more comfortable and effective to use. As one participant noted:
\begin{quote}
\textit{``When one input method is unavailable, I might have multiple alternatives, such as nodding or using voice commands. It would be even better if it could be customized to match my preferences.''} --- P10
\end{quote}

\noindent\textbf{Raising Awareness of SIIDs.}
One interesting finding from our study is that before experimenting \systemname, many participants had not recognized the extent of the situational impairments affecting their daily lives. As a result, they would often deem many tasks unfeasible, give up on them, or seek alternative methods. P4's comment illustrates this realization:

\begin{quote}
\textit{``Before it's more like, there's no way. I have to finish my hand washing and drying it out before I'm able to respond. It's kinda impossible without the system. And the system gives me some new capabilities.''} --- P4
\end{quote}

For the participants, \systemname not only facilitated interaction with their devices but also served as an awareness tool, spotlighting the situational impairments they routinely encountered. This newfound awareness allowed users to extend their capabilities beyond their initial expectations, opening up new possibilities for interactions that they had not previously considered.

\rr{
The insights from our user study involving 10 participants demonstrated the practical application and potential of Human I/O in enhancing user experience in the presence of SIIDs. Future research should explore an extensive deployment study that will involve a larger and more diverse group of participants, which can provide a more comprehensive understanding of the system’s utility across different user demographics and contexts in the long-term.
}


\section{Discussion and Future Work}
\label{sec:discussion}

\noindent\textbf{Privacy and Ethics.}
Maintaining privacy and upholding ethical standards are crucial in the design and deployment of SIIDs systems with active cameras and microphones.

We identify three main concerns:
(1) Invasion of Privacy: Egocentric devices can unintentionally infringe on personal privacy by capturing and recording visuals and sounds without explicit consent.
(2) Data Security and Storage: Using cloud servers for real-time machine learning model inferences requires rigorous measures like data anonymization, encryption, or on-device federated learning to forestall potential data breaches.
(3) Inclusion, Bias, and Discrimination: The deployment of cameras, microphones, and LLMs may unintentionally exclude certain demographics or make inferences based on race, gender, or other protected attributes, risking bias and discrimination.

Acknowledging these issues, we conducted user studies of \systemname exclusively in controlled lab settings. It is imperative for subsequent researchers to prioritize privacy, ethical considerations, and the judicious use of technology. 

While our system is exploratory, advancements in the field hint at promising solutions in future. The recent compact LLMs, such as LLAMA 7B and ALPACA, and fine-tuned models could facilitate on-device computations, ensuring data security. Recent developments by technology companies, exemplified by Apple's Vision Pro, indicate a trend towards encryption and data anonymization protocols in everyday wearable devices. Furthermore, as HMD/AR devices gain prevalence, incorporating learnings from existing egocentric vision research is important. Such studies have already pioneered a variety of privacy-preserving techniques that we could incorporate.

\rr{
\noindent\textbf{Dimensions of availability.}
Our study introduces a four-level classification of channel availability, which has shown high accuracy in diverse scenarios. However, this classification could benefit from expansion beyond a single dimension. Future research should aim to develop a more nuanced understanding of channel availability, considering multiple dimensions such as the duration of the task, type of impairment, effort required in freeing up the channel, effort required in resuming the task, and the ability to multitask, etc.

For example, the dimension of duration highlights the variability in the length of time a channel is unavailable. Short-term unavailability, like a quick glance at a notification, may only need temporary adaptations. In contrast, longer durations of unavailability, such as during meetings, might require more substantial changes in user interface design.
Another example is the ability to multitask. This dimension helps recognize situations where multitasking is feasible versus those where it may be disruptive. This understanding can inform the design of systems that better align with users' capabilities and preferences, reducing the cognitive load and enhancing user experience.

Incorporating these multiple dimensions into the measurement of channel availability can provide users with finer controls over how their devices adapt to their changing needs and situations, and also offer developers with a more comprehensive range of options to consider and design for.
}

\noindent\textbf{Incorporating More Sensing Techniques.}
The current implementation of \systemname primarily relies on an egocentric view camera and microphone sensors, which may limit the system's ability to accurately detect certain aspects of user interactions. For instance, it may not be able to determine if a user is brushing their teeth, wearing headphones, or if their fingertips are wet. To enhance the capability and detection accuracy of \systemname, future systems should incorporate additional sensing techniques.

One promising approach involves utilizing gaze tracking, available in many contemporary XR devices, to measure users' attention and gather more contextual information. Moreover, pupil diameter changes can be measured to estimate cognitive load~\cite{lindlbauer2019context}.
While AR glasses or egocentric cameras provide a rich data source, their continuous usage throughout the day may not be desirable or feasible for users. Developing alternative, lower-resolution sensing methods that leverage mobile devices could be a more practical solution. For example, an approach similar to Google's Activity Recognition API~\cite{activityrecognition} could be adapted to infer user availability based on patterns of device usage and motion data.

\noindent\textbf{Dataset.}
Our prototype has demonstrated the potential for predicting channel availability using a few-shot chain-of-thought reasoning prompt. However, to develop a more robust, faster, and accurate model, and to establish a formal benchmark for evaluating situationally-aware systems, we recognize the need for a large-scale, extensive dataset. This dataset should encompass comprehensive features, similar to those found in the Ego4D dataset~\cite{grauman2022ego4d}, while also incorporating lower-resolution features that can be obtained from everyday devices (\eg smartphone IMU).

\noindent\textbf{Toolkit.}
In our \systemname prototype, we have implemented an extensible reasoning framework to infer SIIDs from multimodal sensors. As a proof-of-concept, we used egocentric cameras and microphones as input sources. By open-sourcing \systemname, we hope to provide a toolkit that empowers XR and sensing researchers and developers to create more accessible systems. For example, by incorporating input from thermal cameras, eye-tracking cameras, depth sensing~\cite{Du2020DepthLab}, Inertial Measurement Unit (IMU) signals, Ultra Wide Band (UWB) signals, street views for remote tourism~\cite{Du2019Geollery}, and the state-of-the-art sensing algorithms, \systemname can be further expanded to create a holistic SIID detection framework for developing wearable applications.


\noindent\textbf{Adaptation Strategies.}
While our paper primarily focuses on the detection of SIIDs, a complete self-adaptive system addressing SIIDs would also require the ``adaptation policies'', or ways to adjust the system's behavior given the detected impairments. However, adaptation strategies would require the efforts of a separate, comprehensive research project, which goes beyond the scope of this work. 

An important question to investigate is whether a universal design can be achieved, and if we can develop algorithms to suggest appropriate adjustments based on user needs and contextual factors. 
Establishing robust evaluation frameworks, such as new metrics and user studies, will enable researchers and practitioners to assess the effectiveness of different adaptation approaches and develop systems that better accommodate user needs in various contexts.

We hope that the detection results from our system might serve as a starting point for app developers, aiding them in considering adaptation policies suited to their specific contexts.

\noindent\textbf{A Situationally Aware Network.}
Another interesting direction for future research is to investigate the development of a situationally aware network that connects multiple devices. For example, a person is on a phone call while their spouse tries to dry their hair, and a message appears on the spouse's device. The network, recognizing the ongoing phone call, could send reminders to prevent noise interference. This illustrates the potential of a collaborative system across multiple devices and multiple people that responds intelligently to a network of context.

\section{Conclusion} 

In this paper, we presented a unified approach to detecting SIIDs based on the availability of human input/output channels. We shared insights from a formative study that guided the design of our system, emphasizing the importance of integrating contextual cues and proposing a four-level scale for measuring channel availability. Furthermore, we introduced \systemname, a system that combines egocentric device, multimodal sensing, and large language models to predict channel availability. Our technical evaluation and user study demonstrated the effectiveness of \systemname, and its potential in reducing user effort and improving performance in the presence of SIIDs. By abstracting SIIDs into channel availability, our work offers a step towards comprehensive detection of situational impairments; and based on this first step, we see an exciting future to build a general-purpose toolkit for SIIDs detection and enable developers to address a range of situational impairments in our daily lives. 


\begin{acks}
We would like to extend our thanks to Siyou Pei, Xiuxiu Yuan, Alex Olwal, Eric Turner, and Federico Tombari for providing assistance and/or reviews for the manuscript. We would also like to thank our reviewers for their insightful feedback.
\end{acks}


\bibliographystyle{ACM-Reference-Format}
\bibliography{_HumanIO}

\appendix

\section{A Decision Tree for Determining the Level of Channel Availability}
\begin{figure}[ht]
\centering
\includegraphics[width=0.85\linewidth]{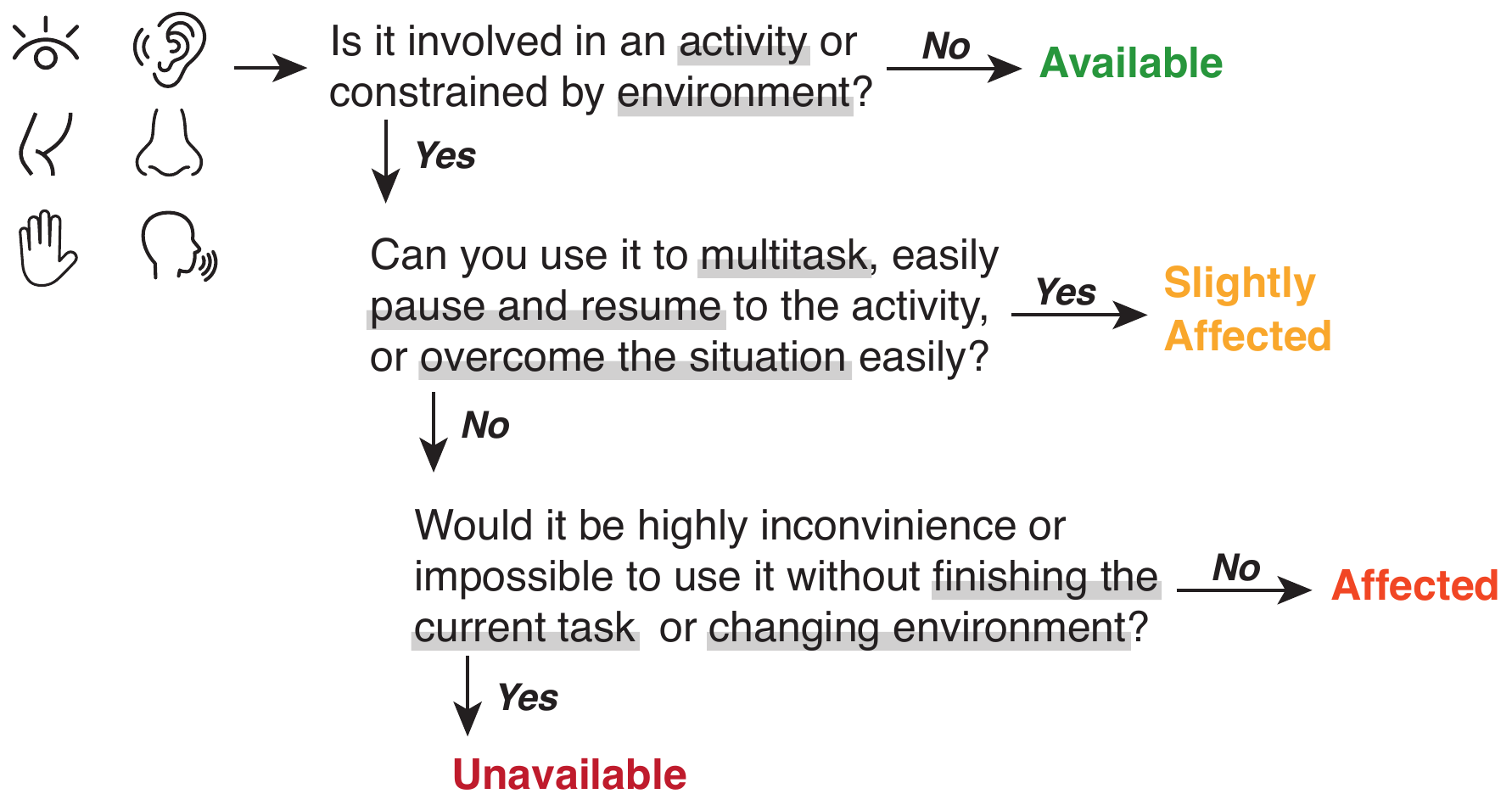}
\caption{A decision tree based on our proposed four-level scale of channel availability.}
\label{fig:decision-tree}
\end{figure}

We sketched a decision tree for determining the level of channel availability based on our proposed scale (\autoref{fig:decision-tree}). We envision our approach may inspire future research to create more holistic and interactive labelling systems for SIIDs. 

\section{Human I/O Implementation Details}
We present implementation details of sensing volume, sound event, and brightness in the \systemname system.

\label{sec:tech-details}

\subsection{Direct Sensing: Object Detection}
To determine if a hand $H$ is holding an object $O$, we apply a rule-based method that considers the following criteria:
\begin{enumerate}
    \item The confidence score of $O$ is greater than 0.70.
    \item The nearest distance between landmarks of $H$ and the bounding box of $O$ is less than 20 pixels.
    \item The average distance between the thumb and the index, middle, ring, and pinky fingers is less than 0.25.
    \item $O$'s predicted label is not ``person''.
\end{enumerate}
These thresholds were determined empirically during system development and have demonstrated effectiveness on various objects. Our method prioritizes the minimization of false positives, resulting in a low recall but higher precision.

\subsection{Direct Sensing: Volume Level}
To sense the direct impact on the availability of the hearing channel, \systemname measures the volume level of the environment. Specifically, we use the Web Audio API to access the user's microphone and the incoming audio stream. 
We maintain a buffer of the 20 most recent volume measurements to provide a continuous and smooth volume estimate. 
The computed volume levels are then converted to decibels: 
\begin{equation}
    \textrm{dB} = 20 \cdot {log_{10}(\textrm{volume})} + 100 
\end{equation}
This allows for a more intuitive representation of the acoustic environment. 
We format the output as: 
\textit{``The environmental volume level is around \texttt{<value>} decibels.''}

\subsection{Direct Sensing: Audio Classification}
To provide a more comprehensive understanding of the audio environment, our system not only measures volume levels but also identifies specific audio events. We leverage YAMNet\cite{yamnet}, a pre-trained deep neural network designed for audio event classification. YAMNet is capable of detecting 521 distinct audio event classes, including barking, laughter, siren, and silence, etc. 
During audio processing, we extracted the input buffer's channel data and fed it to the audio classifier.
We stored the top three category names in an array, and output the top-1 audio class if the confidence score is greater than 0.70. 
We format the output as:
\textit{``The environmental sound may contain \texttt{<audio class>}.''}

\subsection{Direct Sensing: Brightness}
To sense the direct impact on the availability of the vision/eye channel, \systemname measures the brightness of the environment. We utilize relative luminance, a metric accounting for human perception of brightness. We follow the WCAG\footnote{WCAG: \url{https://www.w3.org/WAI/GL/wiki/Relative_luminance}} accessibility guideline to calculate luminance. Specifically, we first convert each frame to sRGB color space and calculate luminance using Rec. 709 coefficients: 
\begin{equation}
    Y = \frac{1}{255}(0.2126 * R + 0.7152 * G + 0.0722 * B)
\end{equation}
We obtain a luminance value ($Y$) for each pixel, and compute the average luminance across all pixels, resulting in a single value representing perceived brightness in the range of 0 to 1.
A value close to 0 indicates a very dark environment, while a value close to 1 indicates a very bright environment. A value around 0.5 suggests a medium brightness level.
Similar to volume level, we applied temporal smoothing by averaging 20 luminance values in the past second.  
We format the output as: 
\textit{``The luminance value of the current environment is \texttt{<value>}, in the range of 0 to 1.''}

\section{Prompts for Large Language Models}
We elaborate on the detailed prompts for reasoning on user activity, their environment, and predicting human input/output channel availabilities.

\label{sec:prompts}
\subsection{Activity}
``An egocentric view of User is showing'' + \\ \texttt{<BLIP-2 output>} + \\ ``Describe what User is doing briefly and objectively, as concisely as possible, without guesses or assumptions. Answer in the format of 'User is...'. If it seems that User is not doing anything, answer `User is not doing anything'. If it cannot be inferred, answer `Unsure'.'' 

\subsection{Environment}
``An egocentric view of User is showing'' + \\ \texttt{<BLIP-2 output>} + \\ ``What location or environment is User likely to be in? Answer in the format of 'User is in...'If it cannot be inferred, answer `Unsure'.'' 

\subsection{Channel Availability Prediction}
\subsubsection{Full Version}
\hfill\\
\textit{[Instruction]}

Available: The channel is currently not involved in any activity, or constrained by any environmental factors. It takes low to zero effort to use the channel to do a new task. Example: A user is sitting at their desk with their hands free, eyes not engaged in any task, and no background noise interfering with their hearing or speech. 

Slightly Affected: The channel is engaged in an activity or constrained by an environmental factor. Given a new task that requires the channel, users can multitask, easily pause and resume to the current activity, or easily overcome the situation. Example: A user is holding a remote control, which can be quickly put down to free up their hand for another task. 

Affected: The channel is involved in an activity or constrained by an environmental factor. Given a new task, the user may experience inconvenience or require effort to use the channel. Example: A user is carrying grocery bags in both hands, making it challenging to use their hands for other tasks without putting the bags down first. 

Unavailable: The channel is completely unavailable due to an activity or environmental factor, and the user cannot use it for a new task without substantial adaptation or changing the environment. Example: A user is attending a loud concert, making it impossible for them to hear incoming notifications or carry on a conversation without stepping outside.
\\

Given the current activity and environment as described below, what are the availability of C’s vision/eye, hearing, vision, and hands/fingers channels? For each channel provide reasons first and then answer using the scale defined above: available, slightly affected, affected  or unavailable. 

Separate each channel prediction with a semicolon (;).
\\

\textit{<Few-shot Example 1>}

Q: C is washing dishes in a kitchen sink. C is in a kitchen. C’s hand is washing dishes. The environmental volume is around 40 dB. 

A: Let’s think step by step. 

Eye Reasoning: While C is washing dishes, their eyes are partially occupied with the task at hand, which involves looking at the dishes to ensure they are clean. However, they can still glance away or multitask to some extent, so their vision is slightly affected but severely impacted.

Eye: Slightly Affected;

Hearing Reasoning: Washing dishes in the kitchen sink does not affect C's hearing. They can still hear other things happening around them.

Hearing: Available;

Vocal Reasoning: C's speech and voice are not impaired by washing dishes. They can still talk or communicate with others while performing the task.

Vocal: Available;

Hand Reasoning: As C's hand is currently engaged in washing dishes, it is not available for other tasks. If C needs to use their hand for something else, they would need to stop washing dishes and wipe their hands.
Hand: Not Available;

[ANSWER COMPLETED]
\\

\textit{<Few-shot Example 2>}

Q: C is playing an acoustic guitar in a room. C is in a small, cozy room with minimal furniture and decorations. C’s hand is playing an acoustic guitar. The environmental volume is around 58 dB.

A: Let’s think step by step. 

Eye Reasoning: C is playing an acoustic guitar, which requires some attention to the placement of fingers on the frets and possibly looking at the sheet music or chords. However, their eyes are not entirely preoccupied with the task and can still be used for other tasks with some level of efficiency, by pausing playing guitar and potential put it away.

Eye: Affected;

Hearing Reasoning: As C is playing an acoustic guitar in a small room, the sound from the guitar is likely to be more noticeable. However, the environmental volume is low, which means that although their hearing may be somewhat affected by the sound of the guitar, they should still be able to hear other things, especially if they are loud or distinct.

Hearing: Affected;

Vocal Reasoning: Playing the guitar does not directly involve using one's voice or speech, so C should be able to use their voice for other tasks while playing the guitar. However, their focus may be divided between playing the guitar and speaking, which could affect their ability to fully concentrate on either task.

Vocal: Available;

Hand Reasoning: C's hand is actively engaged in playing the acoustic guitar. Using their hand for other tasks while playing the guitar would require some effort, as it would require them to stop playing the guitar. 

Hand: Affected;

[ANSWER COMPLETED]
\\

\textit{<Few-shot Example 3>}

Q: C is working at a desk with a laptop. C is in a library. C's hand is typing on a computer. The environmental volume level is around 42 dB. 

A: Let’s think step by step. 

Eye Reasoning:

C is currently using their eyes to focus on the laptop screen in front of them. While their attention is primarily on the laptop, they still have the ability to momentarily glance at other visual stimuli in their environment. However, their ability to focus on other tasks requiring visual attention may be somewhat affected.

Eye: Affected;

Hearing Reasoning:

The environmental volume level is low, which means that C is not experiencing any significant auditory impairment. They should be able to hear other things happening around them without much difficulty.

Hearing: Available;

Vocal Reasoning:

C is in a library, which typically has rules about maintaining a quiet environment. While their voice is physically available, using it for other tasks may be considered inappropriate or disruptive in this setting. Therefore, their ability to use their speech or voice for other tasks is situationally affected.

Vocal: Affected;

Hand Reasoning:

C is currently using their hands to interact with the laptop, such as typing or using the touchpad. They may be able to briefly use their hands for other tasks, but their ability to focus on other hand-related tasks might be affected while they are engaged with the laptop.

Hand: Affected;

[ANSWER COMPLETED]
\\

\textit{[Current Context (example)]}

Q: C is working on a piece of wood. C is in a workshop or a carpentry studio. C’s hand is cutting a piece of wood. The environmental volume level is loud.

A: Let’s think step by step.

\subsubsection{Lite Version}
In the lite version, we remove all reasoning steps in the few shot examples, as well as ``let's think step by step.'':

\textit{[Few-shot Example 1]}

Q: C is washing dishes in a kitchen sink. C is in a kitchen. C hand is washing dishes. The environmental volume is around 40 dB. 

A: 

Eye: Slightly Affected;

Hearing: Available;

Vocal: Available;

Hand: Not Available;

[ANSWER COMPLETED]
\\

\textit{[Few-shot Example 2]}

Q: C is playing an acoustic guitar in a room. C is in a small, cozy room with minimal furniture and decorations. C’s hand is playing an acoustic guitar. The environmental volume is around 58 dB.

A: Eye: Affected;

Hearing: Affected;

Vocal: Available;

Hand: Affected;

[ANSWER COMPLETED]
\\

\textit{[Few-shot Example 3]}

Q: C is working at a desk with a laptop. C is in a workspace or office environment. C's hand is typing on a keyboard. The environmental volume level is around 42 dB. 

A: Eye: Affected;

Hearing: Available;

Vocal: Available;

Hand: Affected;

[ANSWER COMPLETED]

\begin{figure}[ht]
\centering
\includegraphics[width=0.9\linewidth]{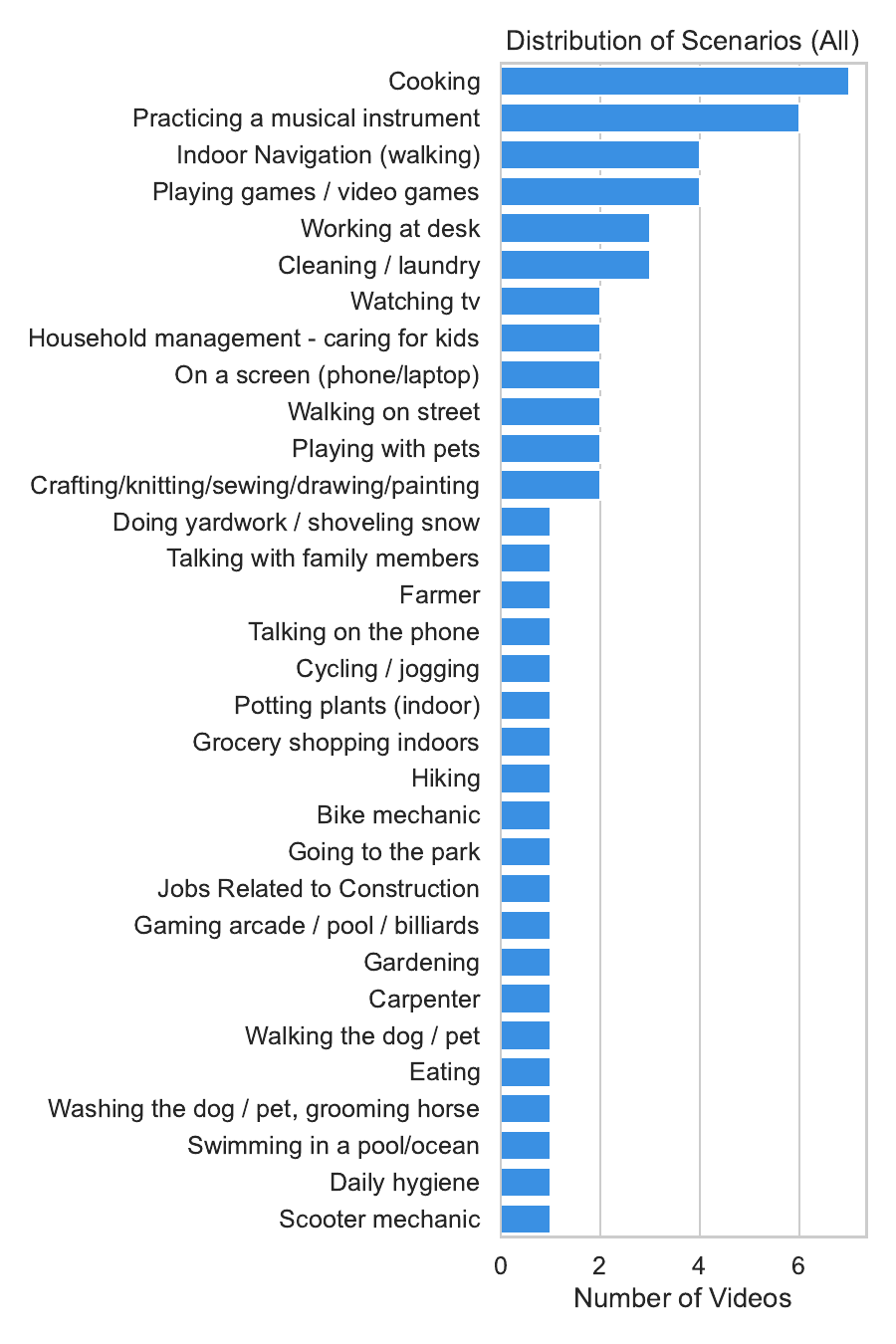}
\caption{Distribution of the scenarios in our evaluation dataset that contains 60 egocentric video recordings from Ego4D dataset~\cite{grauman2022ego4d}.}
\label{fig:scenarios_all}
\end{figure}

\section{Evaluations}

\subsection{Technical Evaluation}
\label{sec:appendix-technical-eval}
\subsubsection{Distribution of Scenarios}

\autoref{fig:scenarios_all} the full distribution of the scenarios of the video recordings used in our technical evaluation.

\subsection{User Study}
\subsubsection{Task Load Index Questions}

We list the set of task load index questions used in Section~\ref{sec:user-study}.
After each task, we ask (from 1-very low, to 7-very high):
\begin{enumerate}
    \item Mental Demand: How much mental and perceptual activity was required? Was the task easy or demanding, simple or complex?
    \item Physical Demand: How much physical activity was required? Was the task easy or demanding, slack or strenuous?
    \item Temporal Demand: How much time pressure did you feel due to the pace at which the tasks or task elements occurred? Was the pace slow or rapid?
    \item Effort: How hard did you have to work (mentally and physically) to accomplish your level of performance?
    \item Frustration Level: How irritated, stressed, and annoyed versus content, relaxed, and complacent did you feel during the task?
    \item Overall Performance: How successful were you in performing the task? How satisfied were you with your performance?
\end{enumerate}



\end{document}